\documentclass[%
 reprint,
 superscriptaddress,
 onecolumn,
 amsmath,amssymb,
 aps,longbibliography
]{revtex4-1}
\usepackage{graphicx}
\usepackage{dcolumn}
\usepackage{bm}
\usepackage{hyperref}
\usepackage[mathlines]{lineno}
\usepackage[bottom]{footmisc}
\usepackage{lipsum}
\usepackage[super]{nth}%
\usepackage{color}
\usepackage{hhline}

\begin{document}
\raggedbottom

\title{Wave transport and localization in prime number landscapes}
\author{Luca Dal Negro}
\email{dalnegro@bu.edu}
\affiliation{Department of Electrical and Computer Engineering, Boston University, 8 Saint Mary\textsc{\char13}s Street, Boston, Massachusetts 02215, USA}
\affiliation{Division of Material Science and Engineering, Boston University, 15 Saint Mary\textsc{\char13}s Street, Brookline, Massachusetts 02446, USA}
\affiliation{Department of Physics, Boston University, 590 Commonwealth Avenue, Boston, Massachusetts 02215, USA}
\author{David Taylor Henderson}
\affiliation{Department of Electrical and Computer Engineering, Boston University, 8 Saint Mary\textsc{\char13}s Street, Boston, Massachusetts 02215, USA}
\author{Fabrizio Sgrignuoli}
\affiliation{Department of Electrical and Computer Engineering, Boston University, 8 Saint Mary\textsc{\char13}s Street, Boston, Massachusetts 02215, USA}


\begin{abstract}
In this paper, we study the wave transport and localization properties of novel aperiodic structures that manifest the intrinsic complexity of prime number distributions in imaginary quadratic fields. In particular, we address structure-property relationships and wave scattering through the prime elements of the nine imaginary quadratic fields (i.e., of their associated rings of integers) with class number one, which are unique factorization domains (UFDs). Our theoretical analysis combines the rigorous Green’s matrix solution of the multiple scattering problem with the interdisciplinary methods of spatial statistics and graph theory analysis of point patterns to unveil the relevant structural properties that produce wave localization effects. The onset of a Delocalization-Localization Transition (DLT) is demonstrated by a comprehensive study of the spectral properties of the Green's matrix and the Thouless number as a function of their optical density. Furthermore, we employ Multifractal Detrended Fluctuation Analysis (MDFA) to establish the multifractal scaling of the local density of states in these complex structures and we discover a direct connection between  localization, multifractality, and graph connectivity properties.
Finally, we use a semi-classical approach to demonstrate and characterize the strong coupling regime of quantum emitters embedded in these novel aperiodic environments. Our study provides access to engineering design rules for the fabrication of novel and more efficient classical and quantum sources as well as photonic devices with enhanced light-matter interaction based on the intrinsic structural complexity of prime numbers in algebraic fields.
\end{abstract}
\pacs{Valid PACS appear here}
\maketitle
\section{Introduction}
The spatial localization of light in open scattering environments with a refractive index that randomly fluctuates over the wavelength scale has attracted intense research activities in the last decades thanks to its wealth of mesoscopic physical effects with potential applications to advanced photonics technology \cite{WiersmaReviewDiso,Lagendijk50yers}. Dielectric random media play an important role in a wide range of optical applications such as, for instance, novel light sources and random lasers \cite{wiersma2008Laser,cao1999random}, photonic filters and waveguides \cite{sebbah2006extended}, lens-less imaging systems \cite{Redding,vellekoop2010exploiting}, broadband sensors and spectroscopic devices \cite{boschetti2020spectral}. Due to the crucial role played by wave interference effects in the multiple scattering regime, the optics of disordered dielectric structures shows profound analogies with the transport of electrons in metallic alloys and semiconductors \cite{Sheng,akkermans2007mesoscopic}. As a result, various mesoscopic phenomena known for the electron wave transport in disordered materials, such as the weak localization of light, universal conductance fluctuations, short/long range speckle correlations, and Anderson localization, have found their counterparts in disordered optical materials as well \cite{Lagendijk50yers,Segev_review,akkermans2007mesoscopic,Sheng}. All these phenomena arise from wave interference contributions in the multiple scattering regime which, in the case of Anderson localization, completely break down the traditional (i.e., perturbative) transport picture and bring light propagation to a complete halt \cite{Sheng,akkermans2007mesoscopic}. However, despite a continued research effort, Anderson localization of optical waves remains an evasive phenomenon since it does not occur in open-scattering three-dimensional random media when the vector nature of light is considered \cite{skipetrov2016red}. This is credited to the detrimental effects of the evanescent-field coupling of randomly distributed vector dipole scatterers that confine electromagnetic waves at the sub-wavelength (near-field) scale in dense scattering media \cite{SkipetrovPRL,Bellando}. In fact, the excitation of sub-radiant (dark) near-field modes, or ``proximity resonances", severely limits the occurrence of the interference loops that drive localization effects in such systems \cite{SkipetrovPRL,SgrignuoliDiffusive}.
Moreover, unavoidable structural fluctuations in random media often limit their applicability to optical device engineering \cite{DalNegroReview}. As a result, there is currently a compelling need to create optical media that are structurally complex, yet deterministic, in order to provide  alternative routes for localization and strong light-matter coupling effects without the involvement of any statistical randomness. 

In order to address this fundamental problem, we have recently introduced a novel approach to wave localization science and technology that leverages the structural complexity of prime numbers in quadratic fields \cite{WangPRB,SgrignuoliMF,SgrignuoliRW,TPSE}. Building on our previous work, in this paper we focus on two-dimensional (2D) arrays of scattering dipoles arranged according to the distributions of prime numbers in imaginary quadratic fields (i.e., in the associated rings of integers) and systematically investigate their structural, scattering, and wave transport properties. For conciseness, we refer to these systems as \textit{Complex Prime Arrays} (CPAs) \cite{WangPRB}. Our primary goal is to provide a comprehensive analysis of the relationship between structural and spectral properties that govern the distinctive optical behavior of the entire class of CPAs with unique factorization properties. In particular, we apply the interdisciplinary methods of spatial statistics \cite{illian2008statistical}, spectral graph theory \cite{Newman_2018}, and multifractal scaling analysis \cite{Ihlen,Kantelhardt} to gain information on the geometrical and connectivity properties of the investigated structures that drive their characteristic localization behavior. Using the Green's matrix method, which has
been extensively utilized in the study of the scattering resonances
of open aperiodic media \cite{DalNegrocrystal2016,DalNegrocrystal2019,SgrignuoliVogel,Lagendijk,SkipetrovPRL,Skipetrov2015,SkipetrovPRB,Skipetrov2020}, we  investigate the spectral statistics of scattering resonances by means of extensive numerical calculations of large-scale aperiodic arrays that cannot otherwise be accessed via traditional numerical methods such as Finite
Difference Time Domain (FDTD) or Finite Elements (FEM).
Importantly, the Green's matrix method allows one to obtain full spectral information and access spatial and temporal localization properties from the 
frequency positions and lifetimes (i.e., the inverse of the resonance width) of all the scattering resonances supported by the investigated systems. Based on the analysis of spectral statistics, we demonstrate a clear Delocalization-Localization Transition (DLT) characterized by a drop of the Thouless number below unity and a switch of the level spacing from repulsion to clustering with a power-law scaling as a function of the optical density of the system. 
Finally, we address the Local Density of States (LDOS) and Purcell enhancement in CPAs and demonstrate Rabi splitting of embedded dipole emitters in the strong-coupling regime.
Our results show that CPAs are gapped systems with a multifractal spectrum of localized resonances  resulting in a far-richer scattering and localization behavior compared to both periodic and uniform random structures. Our comprehensive analysis is meant to stimulate the development of novel compact photonic devices with broadband enhancement of light-matter interactions in both the classical and quantum regimes. 

The paper is organized as follows: in section 2, after a concise introduction of the relevant number-theoretic background, we discuss the diffraction and structural properties of CPAs. In section 3 we address the wave transport and localization properties of CPAs through the analysis of the Green’s matrix spectra, level spacing statistics, spectrally-resolved Thouless numbers. In section 4 we investigate the LDOS properties of CPAs and establish their multifractal features by employing the multifractal detrended fluctuations analysis. Section 5 introduces the radiation analysis of CPAs by focusing on Eisenstein prime arrays and demonstrates the Rabi-splitting of embedded quantum emitters in the strong-coupling regime. The last section draws our conclusions. 

\section{Primes in imaginary quadratic fields: structural properties}
\subsection{Algebraic number theory background}
In order to better appreciate our discussion of the structural and optical properties of CPAs, we first introduce the relevant mathematical background and terminology.    
The idea to generalize the familiar properties of rational numbers to algebraic number fields has a long history in mathematics dating back to the pioneering works of Gauss, Kummer, and Dedekind that initiated Algebraic Number Theory (ANT) \cite{LangANT,Goldman,boyer2011history}. Closely related to the theory of representations of integers by binary quadratic forms, quadratic field theory was largely stimulated by the long search for the solution of Fermat's last theorem, whose complete proof was published by Andrew Wiles in 1995  \cite{wiles1995modular}, as well as by Gauss's seminal work on quadratic reciprocity and cyclotomic fields \cite{Goldman,edwards1996fermat}. The vast subject of ANT rapidly grew into a fascinating area of modern number theory with fundamental connections to many other branches of mathematics \cite{Cohen2,Cohen3,Ireland,Stewart,neukirch2013algebraic}. The basic notion of ANT that we leverage in our paper is the concept of a \textit{quadratic field}, which is defined as a degree $2$ extension of the field of rational numbers $\mathbb{Q}$. The previous statement means that, if we consider $d\neq{1}$ to be a square-free integer (called \textit{radicand}), then the quadratic field denoted by ``$\mathbb{Q}$ adjoin $\sqrt{d}$" is the sets of numbers:
\begin{equation}\label{eq1}
{\mathbb{Q}(\sqrt{d})=\{a+b\sqrt{d}\,\,\, |a,b\in\mathbb{Q}\}}
\end{equation}
By definition,  this set of points forms a linear vector space over $\mathbb{Q}$, generated by the basis $\{1,\sqrt{d}\}$ with rational coefficients. The dimension of such a vector space over the field $\mathbb{Q}$ is called the \textit{degree} of the field extension, which is equal to $2$ in the case of quadratic fields. We note that $\mathbb{Q}(\sqrt{d})$ is a subfield of $\mathbb{R}$ that contains all the sums, differences, and quotients of numbers that we can obtain from $\sqrt{d}$, i.e., it is in fact the smallest extension of $\mathbb{Q}$ that contains $\sqrt{d}$. The \textit{conjugate} of the element $\alpha=a+b\sqrt{d}$ is the element $\bar{\alpha}=a-b\sqrt{d}$ and the \textit{norm} in the field is defined as $N(\alpha)=\alpha\bar\alpha=a^{2}-b^{2}d$. The norm is a multiplicative function, therefore the norm of a product of different elements equals the product of their norms. Elements $u$ with norm $N(u)=\pm{1}$ are called \textit{units}. Two elements $\alpha$ and $\beta$ are called \textit{associates} if they are linked by the multiplication with a unit, i.e., if $\alpha=u\beta$ where $u$ is a unit.
Quadratic fields with $d>0$ are called \textit{real quadratic fields} while the ones with $d<0$  \textit{imaginary quadratic fields}. 

In our paper we discuss the optical properties of CPAs obtained from imaginary quadratic fields. 
Since we will be studying  arrays of \textit{prime numbers} and their distinctive aperiodic distributions, the primes are considered in the \textit{rings of quadratic integers} associated to quadratic fields. We remind that \textit{rings} are fundamental algebraic structures consisting of sets of elements equipped with two binary operations that are connected by a distributive law and satisfy properties entirely analogous to those of the familiar addition and multiplication of integers. Therefore, algebraic rings allow one to extend basic arithmetic concepts, such as divisibility or prime factorization, to more general settings \cite{pinter2010book}. 
A central result of number theory \cite{Ireland} provides a convenient characterization for the integer ring $\mathcal{O}_{K}$ associated to the quadratic field $K=\mathbb{Q}(\sqrt{d})$ as follows: if $d\equiv{1}\bmod{4}$, then $\mathcal{O}_{\mathbb{Q}(\sqrt{d})}=\mathbb{Z}[(1+\sqrt{d})/2]=\mathbb{Z}(\tau)$, otherwise
$\mathcal{O}_{\mathbb{Q}(\sqrt{d})}=\mathbb{Z}[\sqrt{d}]$. This characterization also allows one to define a norm in the ring $\mathcal{O}_{\mathbb{Q}(\sqrt{d})}$ as $N(\alpha)=\alpha\bar{\alpha}=(a+b\tau)(a-b\tau)$ where $a\rightarrow{a+1/2}$ and $b\rightarrow{b+1/2}$ if $d\equiv1\bmod{4}$. It can be shown that if $d<0$, the ring of integers $\mathbb{Q}(\sqrt{d})$  has at most six units, which determine the maximum degree of rotational symmetry of the corresponding CPAs.

An important example of a ring of algebraic integers is the one of the \textit{Gaussian integers}, i.e., $\mathcal{O}_{\mathbb{Q}(\sqrt{-1})}=\mathbb{Z}[i]\subset\mathbb{Q}(\sqrt{-1})$, explicitly defined by the set:
\begin{equation}
{\mathbb{Z}[i]=\{a+{j}b\,\,\, |a,b\in\mathbb{Z}\}}
\end{equation}
where $j=\sqrt{-1}$ denotes the imaginary unit. The Gaussian integers were originally introduced by Gauss in relation to the problem of \textit{biquadratic reciprocity}, which establishes a relation between the solutions of the two biquadratic congruences $x^{4}\equiv{q}\bmod{p}$ and $x^{4}\equiv{p}\bmod{q}$ \cite{Goldman,Ireland}. 
Note that the norm of $\alpha$ in this ring is $N(\alpha)=\alpha\bar{\alpha}=a^{2}+b^{2}$ and this coincides with the square of the distance from the origin of the point with coordinate $(a,b)$ in the complex plane. Geometrically, Gaussian integers can be viewed as points on a square lattice, i.e., points in $\mathbb{R}^{2}$ with integer coordinates. It is also easy to see that the units in the ring $\mathbb{Z}[i]$ are $\pm{1}$ and $\pm{j}$. 


The reciprocity law can also be extended to higher-order congruences, leading for example to the cubic reciprocity law that was originally proven  by introducing the algebraic ring $\mathcal{O}_{\mathbb{Q}(\sqrt{-3})}=\mathbb{Z}[\frac{1+\sqrt{-3}}{2}]\subset\mathbb{Q}(\sqrt{-3})$, denoted $\mathbb{Z}[\omega]$ and specified by the set:
\begin{equation}
{\mathbb{Z}[\omega]=\{a+{e^{2\pi{j}/3}}b\,\,\, |a,b\in\mathbb{Z}\}}
\end{equation}\index{Eisenstein integers}
This is the ring of  \textit{Eisenstein integers}, which are arranged geometrically in a triangular lattice in the complex plane. Eisenstein integers have the 6 units $\pm{1}$, $(\pm{1}\pm{\sqrt{-3}})/2$. It can also be established that for all other negative values of $d$ there are only two units, namely $\pm{1}$.

In order to introduce and characterize the CPAs investigated in this paper we must review some basic facts on the factorization problem in quadratic rings. 
We say that an element of a quadratic ring of integers $\alpha$ is \textit{irreducible} if it cannot be factorized into the product of two non-units. On the other hand, a non-unit element $p$ of a ring is called a \textit{prime} if,
whenever $p$ divides the product $\beta\gamma$ of two elements of the ring, it also divides either $\beta$ or $\gamma$. We remark that the irreducible elements should not be confused with the prime elements. In fact, a prime in a general ring must be irreducible,
but an irreducible is not necessarily a prime. The two concepts coincide only for the special rings that admit a unique factorization. Moreover, the rings that admits a Euclidean algorithm for the greatest common divisor of two elements with respect to their norm are called \textit{Euclidean domains} (or norm-Euclidean).
The concepts reviewed here allow one to naturally generalize many known arithmetical facts from rational integers to algebraic rings. In particular, Gauss was able to prove that unique prime factorization also occurs in the Gaussian and Eisenstein integers \cite{Goldman}. However, rings associated to quadratic fields $\mathbb{Q}(\sqrt{d})$ generally lack the unique factorization property, i.e. the fundamental theorem of arithmetic may fail in generic rings of integers. A classical example is provided by the integer ring associated to the field $\mathbb{Q}(\sqrt{-5})$, which comprises all the complex numbers of the form $a+jb\sqrt{5}$ where $a$ and $b$ are integers. In this set there is no unique factorization because, for instance, the number $6$ can be written as a product of irreducible elements of the ring in more than one way i.e., $6=3\cdot2=(1+\sqrt{-5})\cdot(1-\sqrt{-5})$.
A ring in which every non-prime element can be \textit{uniquely} factored into the product of primes or irreducible elements, up to reordering and multiplication with associates, is called a \textit{Unique Factorization Domain} (UFD). Remarkably, it is possible to show that there are only nine imaginary ($d<0$)  quadratic rings that also are UFDs. These are the ones with radicand $d=-1,-2,-3,-7,-11,-19,-43,-67\textrm{ and}-163$ \cite{Ireland}. This important result, already conjectured by Gauss, is known as the \textit{Baker-Stark-Heegner theorem} and the nine values of $d$ listed above are called \textit{Heegner numbers}. The first five Heegner numbers identify rings that are Euclidean domains. On the other hand, the rings of integers associated to the imaginary quadratic fields with $d=-19,-43,-67,-163$ are \textit{non Euclidean}. This fundamental classification of algebraic rings gives rise to markedly different geometrical and wave localization properties for the corresponding CPAs, which will be addressed in detail in the following sections.  

\begin{figure*}[t!]
\begin{center}
\includegraphics[width=0.7\textwidth]{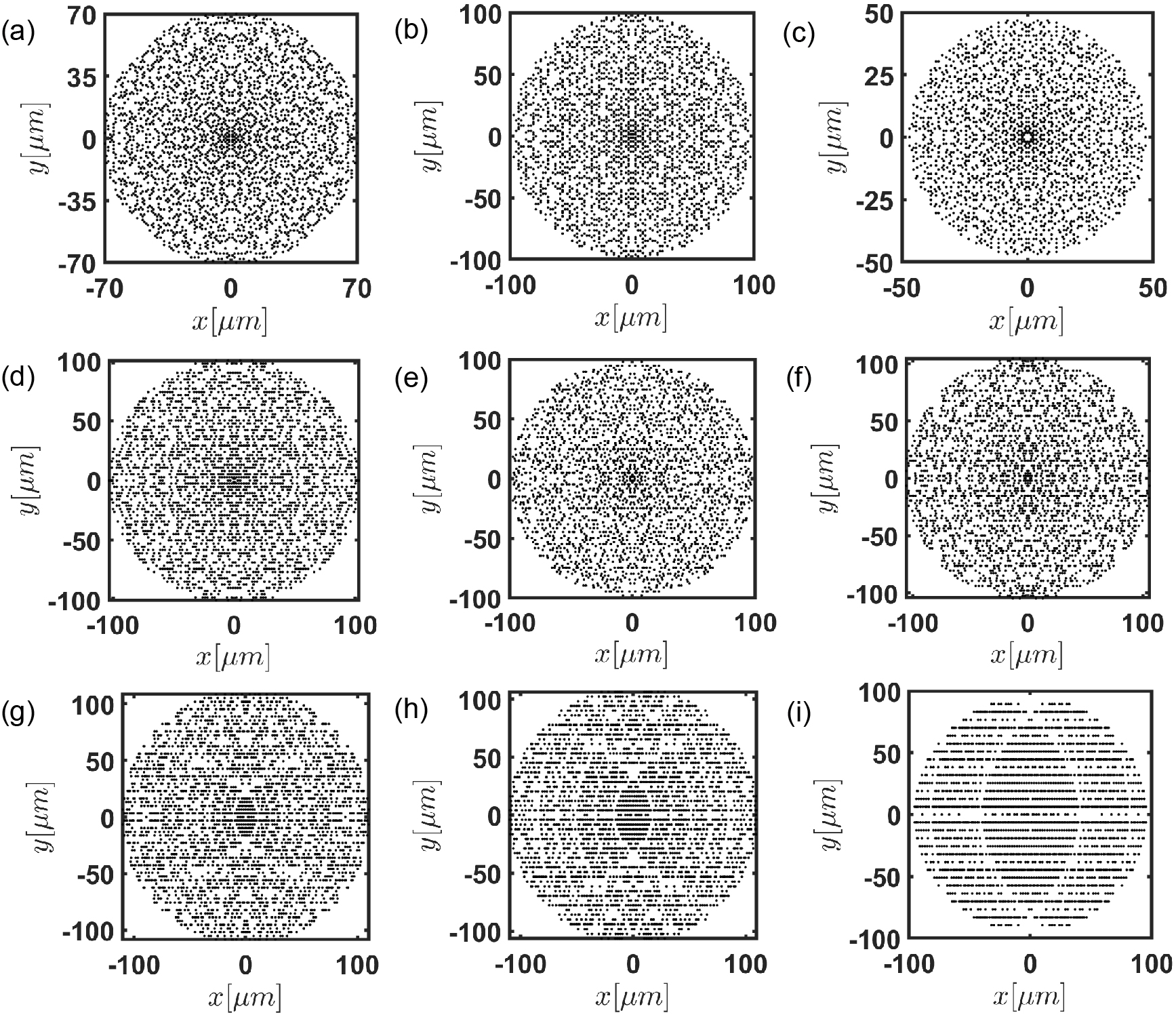}
\end{center}
\caption{Panels (a-i) show the point patterns generated by the prime numbers in the complex quadratic fields $\mathbb{Q}(\sqrt{-1})$, $\mathbb{Q}(\sqrt{-2})$, $\mathbb{Q}(\sqrt{-3})$, $\mathbb{Q}(\sqrt{-7})$, $\mathbb{Q}(\sqrt{-11})$, $\mathbb{Q}(\sqrt{-19})$, $\mathbb{Q}(\sqrt{-43})$, $\mathbb{Q}(\sqrt{-67})$, and $\mathbb{Q}(\sqrt{-163})$, respectively. These arrays are characterized by a number of points $N$ equal to 2658, 2402, 2022, 2582, 2454, 2732, 2908, 2940, and 2254, respectively.}
\label{FigPointPattern}
\end{figure*}
\subsection{Generation and structure of complex prime arrays}\label{generation}
The scattering arrays that we study in this paper correspond to the primes and irreducible elements in the nine imaginary quadratic rings that are also UFDs.
To generate these CPAs, we rely on the fundamental classifications of primes based on the norm and the discriminant of the field \cite{dekker1994prime, prasad2014walks}. The \textit{discriminant} $\delta$ of the imaginary quadratic field $\mathbb{Q}(\sqrt{-d})$ is defined as \cite{prasad2014walks,dekker1994prime}:
\begin{equation}
    \delta=
    \begin{cases}
    d\quad\textrm{if $d\equiv1$ (mod 4)}\\
    4d\quad\textrm{otherwise}
    \end{cases}
\end{equation}
The prime numbers in the ring of integers $\mathcal{O}_{\mathbb{Q}(\sqrt{-d})}$ are found by implementing the three following primality criteria \cite{dekker1994prime,prasad2014walks}:
\begin{itemize}
\item If $N(\alpha)$ is a rational prime, then $\alpha$ is a prime,
\item An odd rational prime $p$ is a prime in $\mathcal{O}_{\mathbb{Q}(\sqrt{-d})}$ if $p$ does not divide $\delta$ and $\Big(\frac{\delta}{p}\Big)=-1$,
\item 2 is prime in $\mathcal{O}_{\mathbb{Q}(\sqrt{-d})}$ if $2$ does not divide $\delta$ and $d\equiv5$ (mod 8).
\end{itemize}
where the \textit{Legendre symbol} $\Big(\frac{\delta}{p}\Big)$ is equal to $-1$ when $\delta$ is not a quadratic residue (mod $p$) \cite{dekker1994prime,prasad2014walks}. With the above criteria, each element in $\mathcal{O}_{\mathbb{Q}(\sqrt{-d})}$ was tested for primality up to a maximum norm value $N_{max}$ that determines the size of the array. Since the norm function for the primes in $\mathcal{O}_{\mathbb{Q}(\sqrt{-d})}$ forms an ellipse, we additionally removed all primes located outside a circular region centered at the origin in order to ensure that all the investigated CPAs are bounded by a circular aperture. 
The generated CPA structures have a comparable number of elements and are shown Figure\,\ref{FigPointPattern}. Specifically, panels (a-i)  display the CPAs corresponding to Heegner numbers $d=-1, -2, -3, -7, -11, -19, -43, -67, -163$, respectively. Moreover, panels (a-e) show the CPAs that correspond to Euclidean fields. All these arrays feature very distinct aperiodic patterns with a characteristic interplay between structure and disorder \cite{tao2005dichotomy}. Moreover, since multiplication by a unit and complex conjugation both preserve primality, these CPAs have different symmetries. Specifically, the Gaussian and Eisenstein primes, shown in panels (a) and (c) respectively, possess 8-fold and 12-fold symmetry while the remaining Euclidean CPA structures have 4-fold symmetry \cite{prasad2014walks,dekker1994prime}.
The coexistence between local structure and global lack of periodicity in these systems is characteristic of the fundamental dichotomy between structure and randomness in the distribution
of primes \cite{tao2005dichotomy}. In particular, Terence Tao showed in
2005 that Gaussian primes contain local clusters of primes with minimal distances, or \textit{constellations}, of any given shape \cite{tao2006gaussian}.
As we will establish in later sections, the structural properties of CPAs have a strong influence on their wave transport and localization properties. In addition, of particular interest for their transport properties are theorems on their asymptotic density $\bar{\rho}=N/\pi R^2$, defined as the number of primes $N$ inside a disk of radius $R$ (i.e., the observation window) in the limit of infinite large systems \cite{prasad2014walks}. 
Note that prime number arrays give rise to \textit{non-homogeneous point patterns} with a density that depends on the size of the observation window. This fact was already known to the young Gauss who, based on numerical evidence, conjectured at age sixteen that the familiar primes (i.e., the primes in $\mathbb{Z}$) are distributed with the non-uniform asymptotic density $\bar{\rho}(x)\sim{1/\ln{x}}$ for $x\rightarrow\infty$.
This observation led to the celebrated \textit{prime number theorem} (PNT), demonstrated in 1896 by Jacques Hadamard and La Vall\'{e}e-Poussin \cite{Goldman}.

Returning to the distributions of complex prime arrays, the asymptotic density is known to be larger for the Eisenstein and Gaussian primes compared to the primes in all the other quadratic fields \cite{prasad2014walks}. We observe that this asymptotic property is also satisfied by the finite-size CPAs that we considered in Figure\,\ref{FigPointPattern}, indicating that the analyzed structures are sufficiently large to manifest the distinctive prime behavior.
In particular, the values of prime densities estimated the analyzed CPAs are $\tilde{\rho}=$ 0.1678, 0.0781, 0.2859, 0.0775, 0.0790, 0.0765, 0.0746, 0.0767, and 0.0783 corresponding to  $d=-1,-2,-3,-7,-11,-19,-43,-67$, and $-163$, respectively. 
As expected, Eisenstein and Gaussian primes feature a larger density for roughly the same number of points, while all the other CPAs have very similar density values. The asymptotic density of primes in the imaginary quadratic fields with $d=-1,-2,-3$
have been established theoretically and satisfy \cite{prasad2014walks}:
\begin{equation}
\frac{\sqrt{2}}{\pi\log{R}}<\frac{2}{\pi\log{R}}<\frac{3\sqrt{3}}{2\pi\log{R}} \end{equation}
where the case $d=-2$ has the smallest density and the Eisenstein primes ($d=-3$) has the largest one given a maximum norm $R$. As it will appear later in this paper, the high degree of rotational symmetry (6-fold) combined with the larger density of the Eisenstein primes makes this CPA platform particularly attractive for the engineering of photonic sources and quantum emitters with high quality factors.  

In panels (f-i) of  Figure\,\ref{FigPointPattern} we show the CPAs that correspond to the non Euclidean imaginary quadratic fields ($d=-19,-43,-67,-163$). 
Differently from the case of the Euclidean structures discussed above, it is evident that all the non Euclidean CPAs share a remarkable structural similarity. In particular, we observe two opposing elliptical shaped patterns of prime gaps resulting from the presence of linear sequences of primes arranged along consecutive horizontal lines. 
Interestingly, the presence of characteristic linear substructures of primes in the non Euclidean CPAs manifests and surprising connection between imaginary quadratic fields and prime generating polynomials that is rigorously established by the \textit{Rabinovitch theorem} discussed in references \cite{goldfeld1985gauss,ayoub1981euler}. The theorem states that for $d<0$ and $d\equiv{1}\bmod(4)$ the following polynomial:
\begin{equation} \label{Rab}
    x^2-x+\frac{1+|d|}{4}=p,\quad x=1,2,...,\frac{|d|-3}{4},
\end{equation}
generates different prime numbers $p$ for the integer values of $x$ shown above if and only if $\mathbb{Q}(\sqrt{-d})$ is a UFD. Therefore, these polynomially generated primes form long horizontal lines (with maximum length proportional to $d$) in non Euclidean CPAs.     
Moreover, these ``lines of special primes" can be found both above and below the origin depending if we consider the polynomial in eq.\,(\ref{Rab}) or its counterpart with a  positive linear term  \cite{ayoub1981euler}. Due to the symmetry of the CPAs, we observe that the number of primes along these lines is equal to $(d-3)/2$, which is twice the number of primes generated by the polynomials. For example, in the CPA with $d=-163$ there are 40 integer values of $x$ such that the above polynomial yields a prime. Consistently, since $\mathbb{Q}(\sqrt{-163})$ is a UFD, in the first quadrant of the corresponding CPA we see a streak of primes (directly above the horizontal line through the center) consisting of 40 elements and there is a similar line directly below it. Clearly, the number of primes doubles to 80 elements if we include also the contributions from the second quadrant. 
Interestingly, similar substructures of primes along remarkable lines determined by quadratic polynomials are also present in other two-dimensional prime arrangements, most notably in the Ulam spiral constructed by writing the positive integers in a square spiral and highlighting only the prime numbers \cite{gardner1964remarkable,stein1967observation}. The symmetry and local structure of CPAs play a central role in the understanding of light transport and localization properties through these complex aperiodic systems. In order to establish robust structure-property relationships in the single scattering regime, we investigate in the following subsections the diffraction properties of CPAs and perform a comprehensive structural analysis based on the associated Delaunay triangulation graphs.

\begin{figure*}[t!]
\begin{center}
\includegraphics[width=0.7\textwidth]{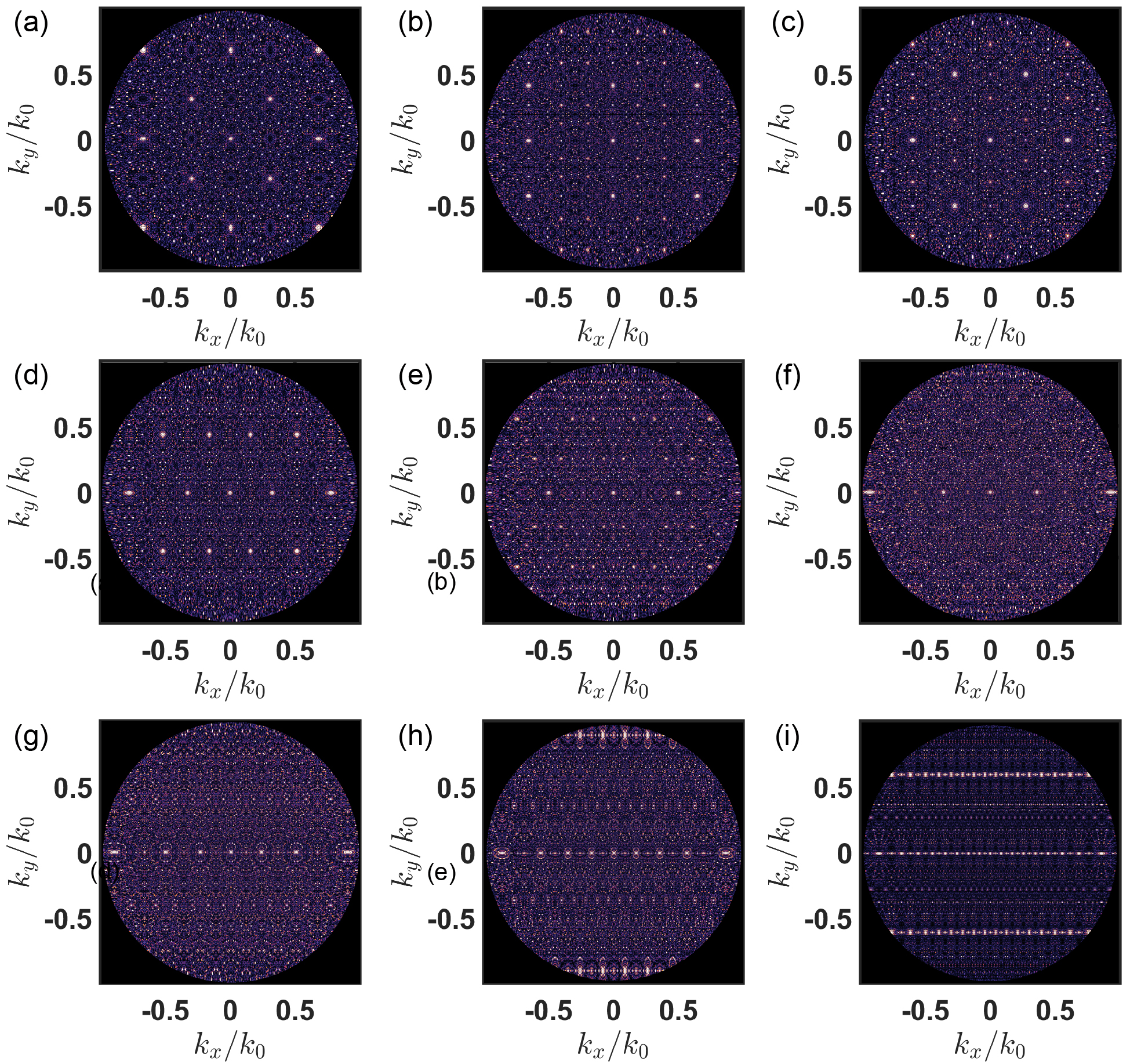}
\end{center}
\caption{Calculated structure factors corresponding to the prime arrays shown in Figure\,\ref{FigPointPattern}. These diffraction patterns are visualized  using the 5th root of the intensity to enhance the contrast.}
\label{FigStructFact}
\end{figure*}
\subsection{Diffraction properties of complex primes}
Aperiodic structures are classified according to nature of their Fourier spectral properties. In fact, the wave intensity diffracted in the far-field can always be regarded as a positive-definite spectral measure and the  \textit{Lebesgue decomposition theorem} provides a rigorous tool for spectral classification. 
This theorem states that any positive spectral measure can be uniquely expressed in terms of three fundamental spectral components: a pure-point component consisting of discrete and sharp diffraction peaks (i.e., Bragg peaks), an absolutely-continuous component characterized by a continuous and differentiable function (i.e., a diffuse background), and a more complex singular-continuous component where the scattering peaks cluster into self-similar structures creating a highly-structured spectral background. 
Singular-continuous spectra are difficult to characterize mathematically but are often associated to the presence of multifractal phenomena \cite{BaakeDiffraction}. Recently, spectral and spatial multifractality has been predicted based on numerical computations \cite{WangPRB} and experimentally demonstrated for Gaussian and Eisenstein prime arrays \cite{SgrignuoliMF,SgrignuoliRW}. In this paper we extend these results to the entire class of CPAs with unique factorization properties.  

When arrays of sub-wavelength scattering particles are illuminated by coherent radiation at normal incidence, the resulting far-field diffraction pattern is proportional
to the \textit{structure factor} defined by \cite{Goldman}:
\begin{equation}
    S_N(\textbf{k})=\frac{1}{N}\sum_{n=1}^{N}\sum_{m=1}^Ne^{-j\textbf{k}\cdot(\textbf{r}_n-\textbf{r}_m)}
\end{equation}
where $\textbf{k}$ is the in-plane component of the scattering wavevector and $\textbf{r}_{n}$ denotes the
vector position of the $N$ particles in the array. Being proportional to the Fourier spectrum of the analyzed point patterns, the structure factor provides access to the spectral classification of aperiodic structures \cite{baake2013aperiodic,dal2013optics,DalNegrocrystal2016}.

In Figure\,\ref{FigStructFact} we report the structure factors computed for all the investigated CPAs. The presence of sharp diffraction peaks embedded in a weaker and highly structured diffuse background, particularly evident in the case of Gaussian and Eisenstein primes, can be clearly observed for all the investigated CPAs. The singular-continuous nature of their diffraction spectra is further supported by the monotonic behavior of the integrated intensity function (IIF) of the local density of states shown in Figure\,S3 of the Supplementary Materials. In particular, we find that the IIF is an aperiodic staircase function with varying slopes that smoothly connect each plateau region due to the emergence of continuous components at multiple length scales \cite{SgrignuoliMF,WangPRB}.
This coexistence of sharp diffraction peaks with a structured background is typical of singular-continuous spectra that are mathematically described by singular Riesz product functions that oscillate at every length scale \cite{baake2013aperiodic}.  Singular-continuous spectra are often found in complex systems with fractal structures, chaotic dynamics, and are also commonly found in traditional quasicrystals \cite{WangPRB}. Singular-continuous spectral components have also been linked to multifractal systems \cite{SgrignuoliMF}. In section \ref{multi} we provide additional evidence of multifractality through the analysis of the density of states and Purcell spectrum of the investigated CPAs. The structure factors shown in Figure\,\ref{FigStructFact} also demonstrate the distinctive symmetries of the CPAs. In particular, we observe 8-fold and 12-fold symmetry in panels (a) and (c) while the remaining structures exhibit 4-fold symmetry. We also remark that the structure factors of the Euclidean CPAs shown in panels (a-e) do not share any common structural motif or pattern due in large part to the significant mismatch of rotational symmetry. 
In contrast, the structure factors of the non Euclidean CPAs, which are shown in panels (f-i),  display common features characterized by the presence of bright peaks along the horizontal lines. Moreover, the number of sharp diffraction peaks increases as the radicand $d$ of the field increases as well. Interestingly, we found that the number of peaks along the horizontal lines is equal to $(d+41)/12$,  reflecting the connection with the previously discussed linear patterns of polynomially generated primes. 
\subsection{Spacing analysis of complex primes}
Traditional spectral Fourier analysis alone is unsuitable for the characterization of the local structural features of aperiodic point patterns and must be complemented by the interdisciplinary methods of spatial statistics \cite{illian2008statistical,DalNegrocrystal2016}. 
\begin{figure*}[t!]
\begin{center}
\includegraphics[width=0.69\textwidth]{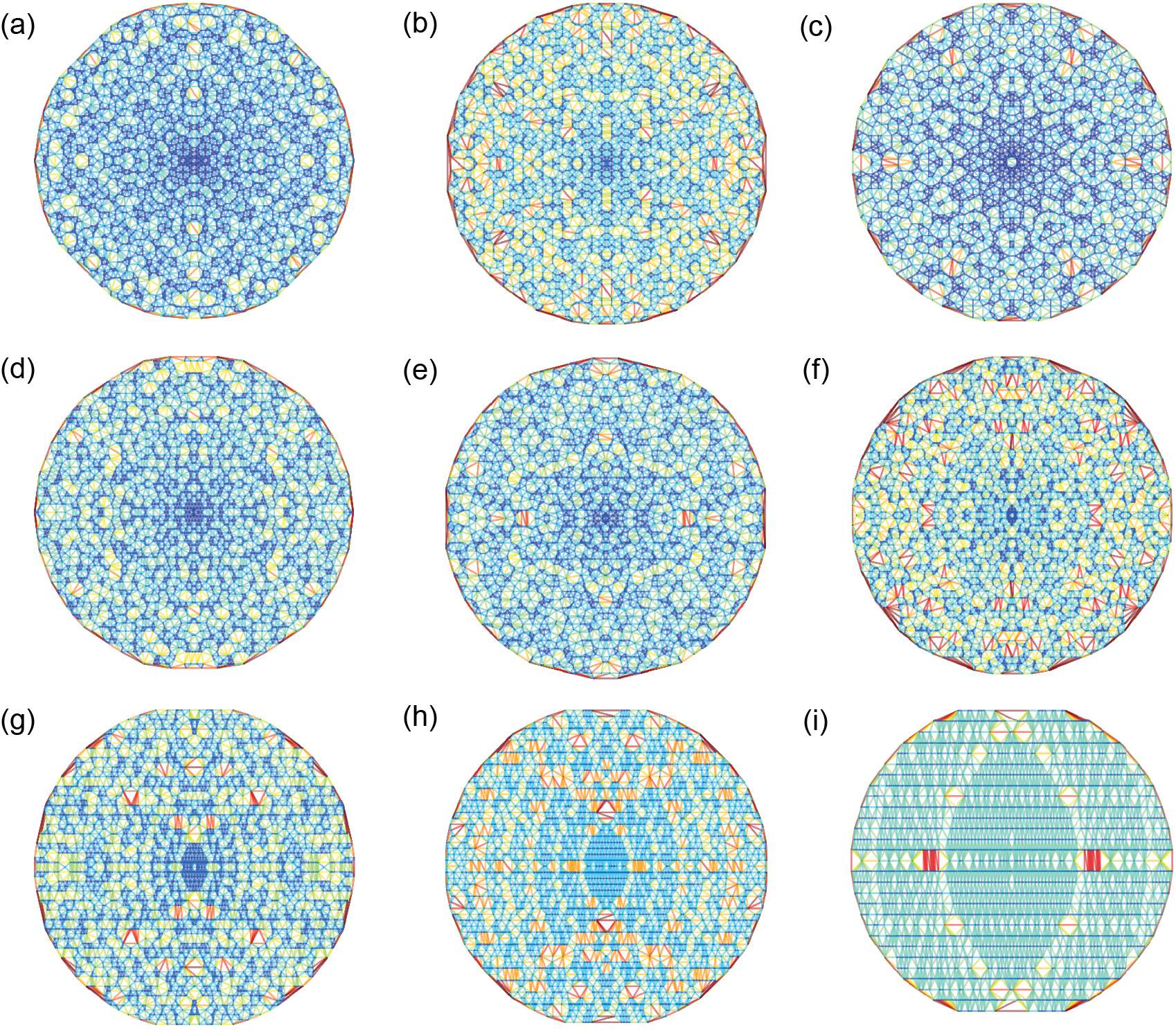}
\end{center}
\caption{(a-i) Delaunay triangulation with color-coded edge according to their lengths of the point patterns displayed in Figure\,\ref{FigPointPattern}. The edge lengths increase from blue (i.e., the shortest edge of the Delaunay triangulation of each CPA) to red (i.e., the longest edge of the Delaunay triangulation of each CPA).}
\label{FigDelaunay}
\end{figure*}
In particular, in what follows we will focus on graph theory methods to further investigate the structure and connectivity properties of CPAs. 
In order to better understand the unique geometrical features of the CPAs, it is instructive to first perform spatial Delaunay triangulation analysis using  color-coded edge lengths in order to visualize the distribution of distances between neighboring primes in the different structures \cite{DalNegrocrystal2016,Trevino}.
The Delaunay triangulation provides a convenient way to impose a graph structure onto the CPAs and this approach can also be employed to study the diffusive transport through arbitrary point patterns \cite{Newman_2018,prasad2014walks}. In Figure\,\ref{FigDelaunay} we display the color-coded Delaunay triangulations for all the investigated CPAs. The edge length  is color-coded with increasing numerical values from blue to red color. In panels (a) and (c) we  observe that the CPAs of the fields $\mathbb{Q}(\sqrt{-1})$ and $\mathbb{Q}(\sqrt{-3})$ support multiple paths from the center to their outer boundaries comprised solely of short edges. In contrast, the CPAs analyzed in panels (b) and (d-i) feature a broader distribution of edge lengths. The color-coded Delaunay can be used to identify areas of similar connectivity lengths inside the overall structure \cite{DalNegrocrystal2016}. In the investigated CPAs, short edges are predominantly localized near the center regions. This tendency is most prominent inside the elliptical regions surrounding the centers of the non Euclidean CPAs shown in panels (f-i). We also observe from Figure\,\ref{FigDelaunay} that there is no smooth variation in edge lengths, but rather they only take on specific values that are localized within specific areas of the structures. This is demonstrated by the repetitions of the same colors within the structures. 
\begin{figure*}[t!]
\begin{center}
\includegraphics[width=0.7\textwidth]{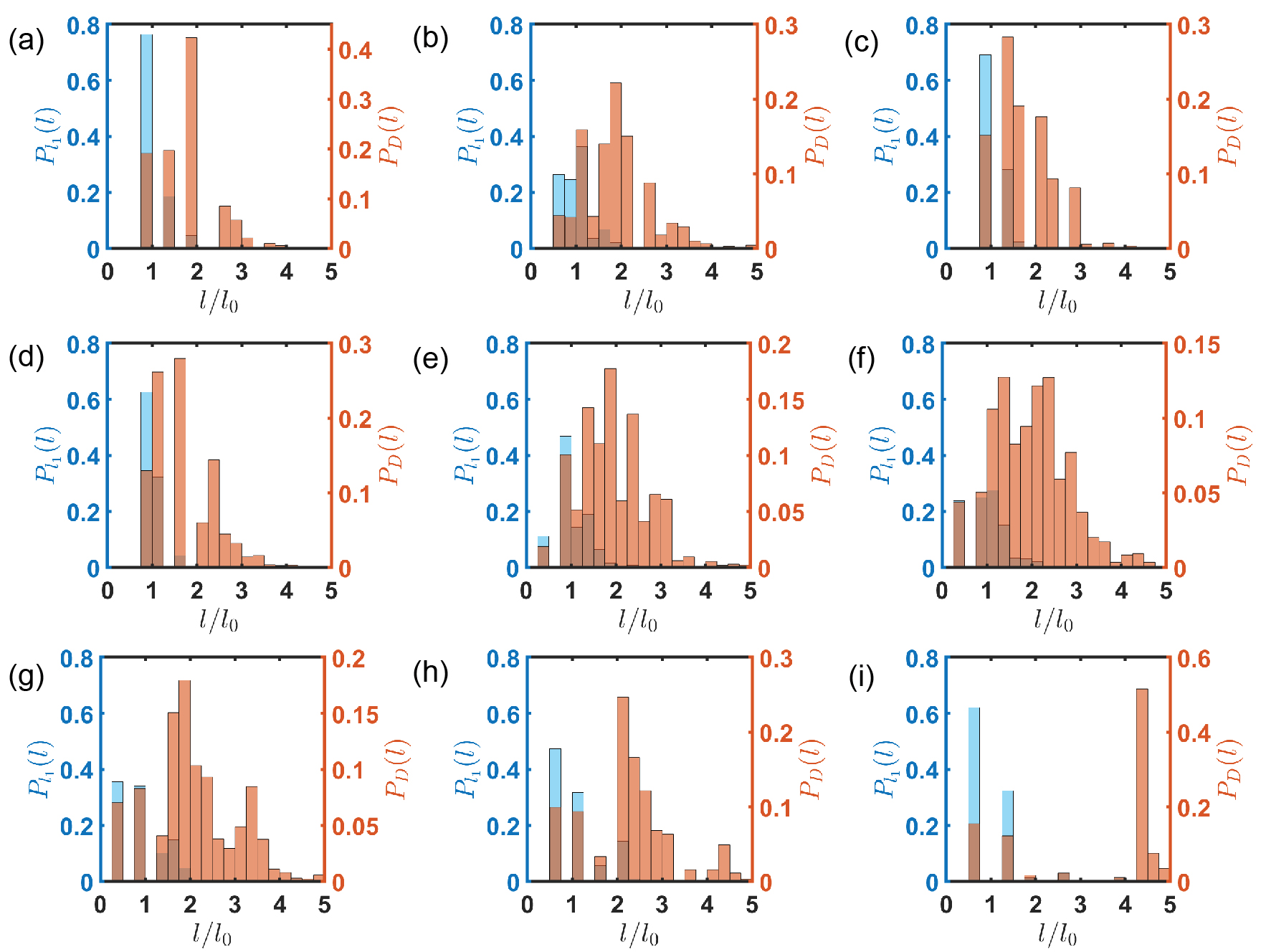}
\end{center}
\caption{Panels (a-i) show the probability density function $P_{d_1}$ of the nearest-neighbor distances of the point patterns reported in Figure\,\ref{FigPointPattern} together with the probability density function $P_D$ of the Delaunay triangulation edge lengths shown in Figure\,\ref{FigDelaunay}. These data are reported as a function of $l/l_0$, where $l_0$ is the mean nearest neighbor distance in each configuration.}
\label{FigEdge}
\end{figure*}
We then study the distribution of the spacing of primes in the CPAs by computing the statistical distributions of the edge lengths of the Delaunay triangulations and by comparing them with the nearest-neighbor distances obtained from the corresponding point patterns. Our results are summarized in the histogram plots shown in Figure\,\ref{FigEdge} and confirm that only few discrete values of nearest-neighbor distances and edge lengths can be obtained in all cases. The non Gaussian nature of these edge distributions unveils structural correlations that markedly separate the aperiodic behavior of CPAs from the well-characterized case of Poisson point patterns \cite{illian2008statistical}. Interestingly, the occurrence of only few values of the nearest-neighbor distances in such large-scale aperiodic sequences is also observed in uniform sequences modulo 1, which display a complex quasi-periodic behavior \cite{miller2021invitation,Niederreiter,Kuipers}. In this case, the nearest neighbor spacing can only assume three values \cite{miller2021invitation}.
In order to extend our geometrical approach, we will introduce in the
next section additional metrics that deepen our understanding of the salient geometrical and local connectivity features of the graphs associated with CPAs, with important implications for the transport and localization of light in these complex structures.

\subsection{Graph theory analysis of complex primes}
In this section, we apply  several quantitative metrics from graph theory to better analyze the distinctive correlation and connectivity properties of the investigated CPAs. In particular, geometrical and topological parameters will be constructed from the graph structure of the Delaunay triangulation associated with each point pattern. In Table 1, we report the computed graph-based topological parameters associated with the unweighted and weighted Delaunay triangulation of each CPA. To facilitate understanding, we provide below a brief introduction to all the considered graph parameters.
\begin{table}[b!]
\begin{center}
    \begin{tabular}{c c c c c c c c}
    \hline
    Structures& $\langle D\rangle$&$R$&$C_N$&$C_{WS}$&$AC$&$C^w$& $\langle S_{node}\rangle$\\
    \hline
    $\mathbb{Q}(\sqrt{-1})$&5.9601&-0.15889&0.38463&0.42557&0.007222&0.4253&2.4673\\
    $\mathbb{Q}(\sqrt{-2})$&5.9609&-0.19228&0.38433&0.43316&0.0088351&0.4325&2.4578\\
    $\mathbb{Q}(\sqrt{-3})$&5.9476&-0.16895&0.38493&0.4256&0.010545&0.4246&2.4712\\
    $\mathbb{Q}(\sqrt{-7})$&5.962&-0.16709&0.38531&0.42495&0.0082481&0.4242&2.4731\\
    $\mathbb{Q}(\sqrt{-11})$&5.9617&-0.18754& 0.38547&0.4251&0.009004&0.4246&2.4575\\
    $\mathbb{Q}(\sqrt{-19})$&5.9656&-0.1617&0.38198 &0.42933 &0.0079667&0.4287&2.4379\\
    $\mathbb{Q}(\sqrt{-43})$&5.9512&-0.21432 &0.37683&0.43486&0.0060718&0.4341&2.4135\\
    $\mathbb{Q}(\sqrt{-67})$&5.9544&-0.22991&0.36705&0.44425&0.0052759&0.4439&2.3831\\
    $\mathbb{Q}(\sqrt{-163})$& 5.9441&-0.21767&0.35096&0.45546&0.0040188&0.4553&2.2650\\
    \hline
    \end{tabular}
\end{center}
\end{table}
We first consider the average node degree $\langle D\rangle$ of the CPAs, which is the average number of edges per node. We can obtain $\langle D\rangle$ from the diagonal entries of the graph Laplacian matrix \cite{Newman_2018}. Interestingly, we notice that the average node degree does not vary significantly across the different structures
due to the homogeneous nature of the corresponding Delaunay graphs. 
However, small variations can be observed when considering the Pearson correlation coefficient $R$, also known as the
\textit{assortativity coefficient}. This
coefficient measures the spatial correlations between nodes with similar degree values, and it is computed as:
\begin{equation}
    R=\frac{1}{\sigma^2_q}\sum_{k_i,k_j}[e(k_i,k_j)-q(k_i)q(k_j)]
\end{equation}
where $k_j$ is the node degree of the $j$-th node, $q(k_j)=(k_j+1)p(k_j+1)/\sum_ik_ip(k_i)$ and $\sigma_q^2$ is the standard deviation of $q(k_j)$, the node distribution \cite{DalNegrocrystal2016,Newman_2018}. The coefficient $R$ ranges between $-1$ and $1$, with positive values indicating a spatial correlation between nodes of similar degree and negative values indicating a correlation between nodes of different degrees, i.e., where high-degree nodes have a tendency to attach to low-degree ones. This last network property is called \textit{disassortativity}. It is generally the case that large social networks are assortative in contrast to biological networks that typically show disassortative mixing, or disassortativity \cite{Newman_2018,estrada2015first}.
From Table 1, we can see that all the investigated CPAs are disassortative. This behavior stems from the distinctive spatial non-uniformity of the CPAs that has been recently characterized, for Gaussian and Eisenstein primes, using multifractal analysis \cite{SgrignuoliMF,SgrignuoliRW}. In addition, here we find that the CPAs of $\mathbb{Q}(\sqrt{-67})$ feature the strongest disassortativity, and that this property is displayed particularly by non Euclidean CPAs. We also computed the $\textit{Newman global clustering coefficient}$ $C$ of the graphs, which measures to what extent the graph nodes tend to cluster together. Since this metrics is based on cluster triplets, it quantifies the tendency of forming ``isolated islands" in the analyzed graph structures \cite{DalNegrocrystal2016,Newman_2018}. The global clustering coefficient is defined as \cite{DalNegrocrystal2016}:
\begin{equation}
    C_N=\frac{3N_T}{P_2}
\end{equation}
where $N_T=Tr(A^3)/6$ is the number of triangles, $A$ is the graph adjacency matrix, and $P_2$ is the number of paths composed of 2 edges in the graph that can be obtained by the formula $P_2=\sum_{i=1}^Nk_i(k_i-1)/2$. From Table 1, we see that the Euclidean structures have a stronger tendency to cluster together globally compared to their non Euclidean counterparts, as it can also be directly appreciated by inspecting  Figure\,\ref{FigPointPattern}. For example, we notice many isolated ``pockets of primes" in the Eisenstein array shown in Figure\,\ref{FigPointPattern} panel (c). This is in strong contrast with the distribution of primes in the non Euclidean fields that tend to align along horizontal lines, as can be seen in Figure\,\ref{FigPointPattern}\,(f-i). 
It is also interesting to consider the local clustering properties of the graphs that are described by the \textit{Watts-Strogatz index} $C_{WS}=(\sum_{i=1}^{N}C_i)/N$, where $N$ is the total number of nodes and $C_i$ is the local clustering coefficient defined as \cite{Newman_2018}:
\begin{equation}
    C_i=\frac{R_i}{k_i-1}.
\end{equation}
Here, $R_i$ is the $\textit{redundancy}$ of vertex $i$, defined as the mean number of connections from a neighbor of $i$ to the other neighbors of $i$ \cite{DalNegrocrystal2016, Newman_2018}. This quantity measures how close nodes and their neighbors are to forming their own complete graph \cite{Newman_2018}. 
As we can appreciate from Table 1, the non Euclidean structures generally outperform their Euclidean counterparts also with respect to this quantitative metric, as the nodes of these graphs are more likely to form complete subgraphs with their neighboring vertices. This can be observed especially in Figure\,\ref{FigDelaunay}\,(f-i) where the primes located within the central elliptic-shaped regions form complete subgraphs with their neighboring vertices. This explains the increase in $C_{WS}$ as $|d|$ rises, since the number of points in the elliptical-shaped region increases with $|d|$. On the other hand, the algebraic connectivity (AC) of a graph is a measure on how well-connected is the graph overall. It is defined as the second smallest eigenvalue of the graph Laplacian matrix \cite{Newman_2018}. We observe in Table 1 that the AC of non Euclidean structures rapidly decreases as the radicand $d$ of the corresponding quadratic field increases. 

We now consider the analysis of  weighted graphs where the weight of each edge is equal to the  Euclidean distance between the two neighboring vertices. For a given vertex $v_i$, the weighted clustering coefficient $c_i^w$ is defined as \cite{barrat2004architecture,Newman_2018}:
\begin{equation}
    c_i^w=\frac{1}{s_i(k_i-1)}\sum_{j,h}\frac{(w_{ij}+w_{ih})}{2}
\end{equation}
where $w_{ij}$ is the weight on the edge connecting vertices $v_i$ and $v_j$, while $s_i$ is the strength of vertex $v_i$ defined by the summation $\sum_jw_{ij}$ in which $j$ ranges over all adjacent vertices to $v_i$ \cite{barrat2004architecture}. In Table 1, we report the weighted clustering coefficient averaged over all vertices in the graph, which we denote as $C^w$. The small difference observed in Table 1 between the weighted clustering coefficient and the Watts-Strogatz index reflects the fact that the local clustering behavior of the CPAs is independent of the length between edges. Finally, we computed the \textit{vertex entropy} averaged averaged over all the vertices of each CPA graphs. The entropy on vertex $v_i$ is denoted by $S_{v_i}$ and it is defined as \cite{sen2019ranking}:
\begin{equation}
    S_{v_i}=-\sum_{m}q_{i,m}\log_2(q_{i,m})
\end{equation}
where $m$ ranges over all adjacent vertices $v_m$ of the vertex $v_i$ and $q_{i,m}$ is defined as $q=w_{i,m}/\sum_jw_{i,j}$, where $w_{i,j}$ is the weight of the edge connecting vertex $v_i$ and $v_j$ and $j$ ranges over all adjacent vertices of $v_i$. 
We observe a decreasing trend in nodal entropy from the non Euclidean CPAs that correlates well with the previously discovered drop of their AC. The data in Table 1 establish the particular importance of the AC and the nodal entropy as sensitive characterization parameters for the geometrical analysis of CPAs and provide a comprehensive set of metrics for the development of structure-property relationships. By addressing the challenging moat problem, in the next section we will provide an initial discussion of the connection between the transport and the structural properties of CPAs.
\begin{figure}[b!]
\begin{center}
\includegraphics[width=0.5\textwidth]{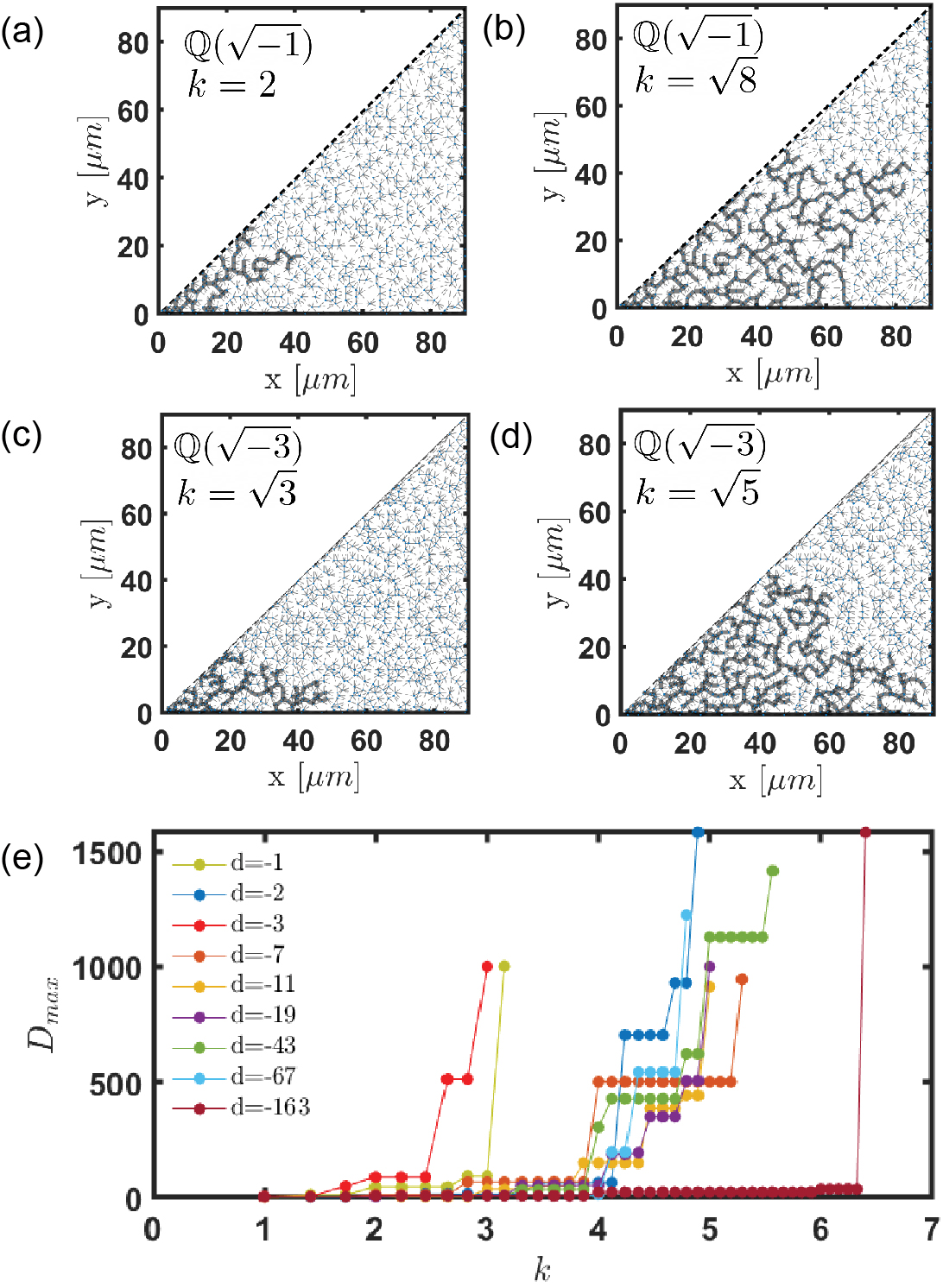}
\end{center}
\caption{Minimum spanning tree analysis applied to the Gaussian prime array when the maximum allowed edge length $k$ (also named moat value) is equal to 2  (panel a) and $\sqrt{8}$ (panel b). Panels (c) and (d) show the same study performed on the Eisenstein configuration when $k$ is equal to $\sqrt{3}$ (panel c) and $\sqrt{5}$ (panel d). Panel (e) compares the maximum distance $D_{max}$, evaluated form the origin of each structures, of all the investigated CPAs for given moat values $k$.}
\label{FigMoat}
\end{figure}

\subsection{Transport and the moat problem}
Diffusive transport through complex systems can be described in a first approximation using the tools of spectral graph theory \cite{Newman_2018,estrada2015first}.
In this simple picture, a scalar quantity diffuses over time through the edges of the graph with a diffusion dynamics determined by the spectral properties of its adjacency matrix (more precisely of its graph Laplacian operator). Therefore, understanding the geometry and connectivity characteristics of the graphs imposed over a complex point pattern can provide invaluable information on the nature of its transport properties, and help identifying robust structure-property relationships. Moreover, the network or graph approach to diffusion through two-dimensional CPAs has a direct connection with one of the unsolved problems in number theory, i.e., the Gaussian moat problem \cite{guy2004unsolved}. A general moat problem asks whether it is possible to walk to infinity by taking steps of bounded length on primes. When considering rational primes (prime numbers in $\mathbb{Z}$) this is clearly not possible since $\{(n+1)!+i\}$ with $i=2\ldots,n+1$ are $n$ consecutive composite integers. In other words, $\limsup\{p_{n+1}-p_{n}\}=\infty$, meaning that there are arbitrarily large gaps in the rational primes. However, the moat problem can be generalized to Gaussian primes or to primes in other quadratic rings, where it is concerned with the characterization of prime-free regions that encircle the origin of the complex plane \cite{prasad2014walks,gethner1998stroll,west2017further,bressoud2000course}.
In particular, the Gaussian moat problem asks whether it is possible to walk to infinity by taking steps of bounded length over the Gaussian primes, which is currently an open question in number theory \cite{guy2004unsolved}. In this context, the existence of a $k$-moat means that it is not possible to walk to infinity with a step size smaller than $k$ \cite{prasad2014walks}. 
The mathematical perspective behind the moat problem is clearly relevant to achieve a rigorous understanding of the transport properties of CPAs based on the relative ease of $k$-moats formation. 
This approach has also been recently applied to the transport through Guassian and Eisenstein primes \cite{prasad2014walks,gethner1998stroll,west2017further}. 

In order to extend this approach to all the considered CPAs, here we use the graph created from the Delaunay triangulation to computationally explore the farthest distances reachable in the CPAs, given a maximum step size. We find that the previously introduced AC and nodal entropy parameters of the CPAs play indeed an important role in this context and largely control the transport properties of CPAs.
To quantitatively study this problem we first create weighted graphs by assigning a weight to each edge equal to the distance between the two primes connected by that edge. We then can find the farthest distance reached for a given moat value $k$ by removing all the edges with  lengths greater than $k$. In the resulting graph, the node associated with the point of largest norm in the Minimum Spanning Tree (MST) is called the frontier prime \cite{prasad2014walks,Newman_2018}. The norm of this prime yields the farthest distance reached with a maximum step size $k$. Since this is a hard computational problem, the symmetry of the structures must be exploited to limit the investigation to only a portion of the structures \cite{prasad2014walks}. Specifically, an octant of the plane is considered for the Gaussian primes, a twelfth of the plane for the Eisenstein primes, and a quadrant of the plane for all the remaining CPAs \cite{prasad2014walks}.  
Representative examples of this method are shown in Figure\,\ref{FigMoat}. In particular, in panels (a) and (b) we show the farthest primes reached when the considered moat value is $k=2$ and $k=\sqrt{8}$, respectively, and the CPA corresponds to the field $\mathbb{Q}(\sqrt{-1})$. It is clear from the plots that a farther prime element and a further distance from the origin are reached when the moat size is increased. Similarly, panels (c) and (d) show the farthest primes reached when the considered moat values are $k=\sqrt{3}$ and $\sqrt{5}$, respectively, for the CPA of $\mathbb{Q}(\sqrt{-3})$. By comparing panels (b) and (d) we can observe that the farthest distance reached is similar in both CPAs even though the maximum allowed step size in the CPA of $\mathbb{Q}(\sqrt{-3})$ is smaller. It follows that the transport through the Eisenstein primes reaches a farthest distance for a given maximum step size compared to the Gaussian primes. This example suggests a method to assess the ease of transport through all the investigated CPAs. Building on this idea, we systematically investigate the farthest distance that is reachable for a given moat value $k$ in all the CPAs with Heegner number discriminant $d$. Figure\,\ref{FigMoat}\,(e) shows the results of our analysis for all the CPAs (labelled in the legend by the values of their discriminant) when considering moat values $k$ ranging from 1 to $\sqrt{41}$. Our results indicate that, given a moat value $k$, the farthest distance will be reached by the Eisenstein primes, while the shortest distance (i.e., most impeded transport) is obtained in the CPA of $\mathbb{Q}(\sqrt{-163})$. Moreover, we found that the transport through the non Euclidean CPAs is generally more impeded compared to their Euclidean counterparts. 
We can qualitatively explain these results based on the difference between the structural and graph connectivity parameters of these two classes of graphs. In particular, as we discussed in the previous section, Euclidean/non Euclidean CPAs differ fundamentally in the values of their AC, density, and node entropy (see Table 1). In particular, the Eisenstein primes have the largest AC, entropy, and density, describing a compact, structurally complex, and highly-interconnected network that favors diffusion processes compared to less dense and more ordered structures with reduced connectivity. In the next section, based on rigorous electromagnetic scattering theory, we will address the optical wave scattering and transport properties of CPAs and establish the validity of structure-property relationships in these complex systems.

\begin{figure*}[t!]
\begin{center}
\includegraphics[width=0.87\textwidth]{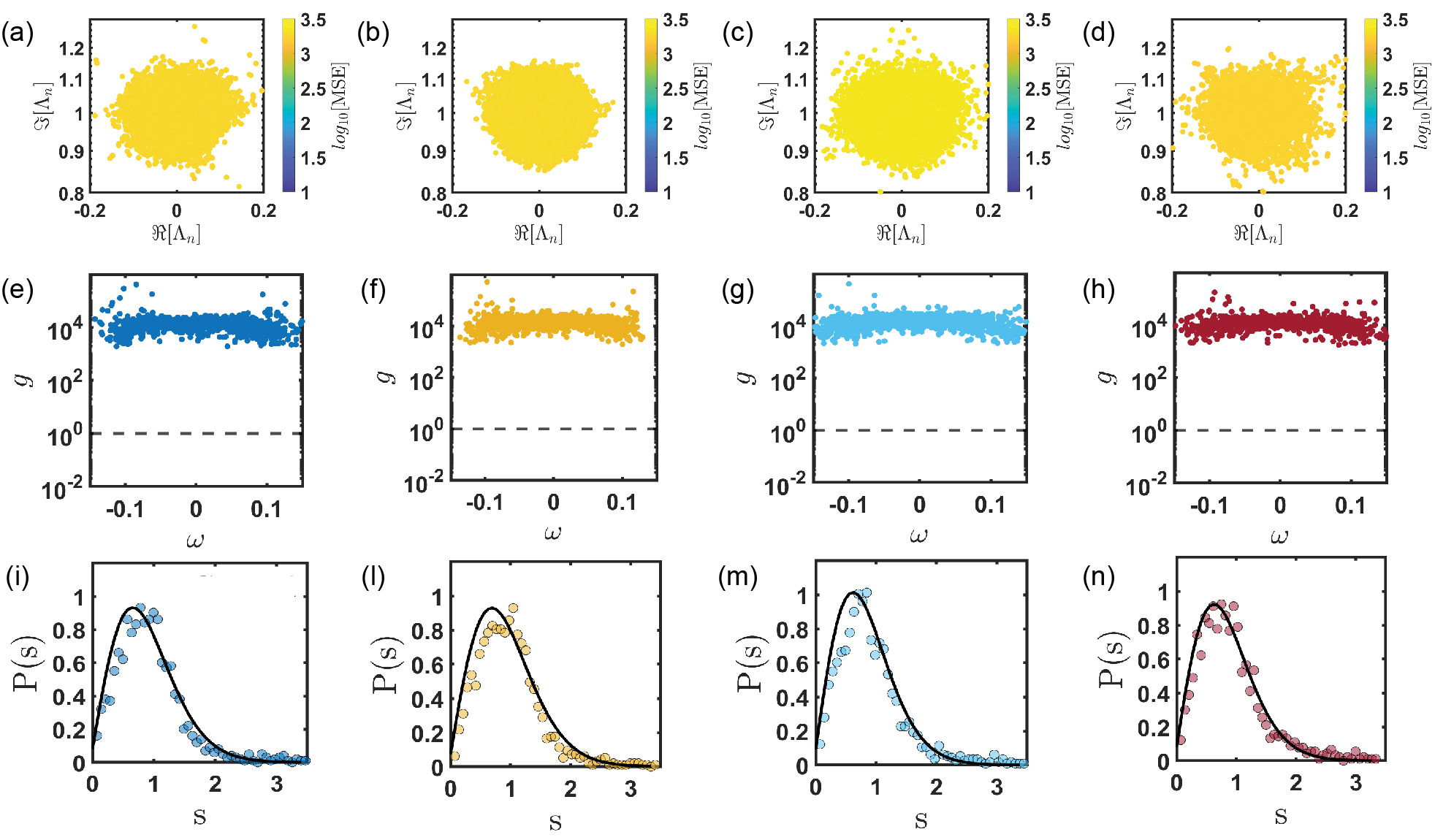}
\end{center}
\caption{Eigenvalues distribution of the Green's matrix\,(\ref{Green}) in the low scattering regime (i.e., $\rho\lambda^2=10^{-6}$) for two representative Euclidean (panels (a-b)) and non Euclidean (panels (c-d)) CPA. In particular, panels (a-d) refer to the arrays generated by following the prime elements in the complex quadratic fields $\mathbb{Q}(\sqrt{-2})$, $\mathbb{Q}(\sqrt{-11})$, $\mathbb{Q}(\sqrt{-67})$, and $\mathbb{Q}(\sqrt{-163})$, respectively. The Thouless number $g$ as a function of the frequency $\omega$ and the level spacing statistics $P(s)$ extrapolated from the distributions of panels (a-d) are reported in panels (e-h) and (i-n). These data demonstrate that the investigated arrays are in the diffusive regime: $g$ is always larger than unity (horizontal black dashed line in panels (e-h)) and $P(s)$ is well-reproduced by the critical statistic in eq.\,(\ref{CriticalDistribution}) (continuous black line in panels (i-n)).}
\label{FigLowOD}
\end{figure*}
\section{Wave transport through prime number landscapes}
We now investigate the spectral and wave transport properties of vertically-polarized electric dipoles spatially arranged according to the distributions of primes in imaginary quadratic fields. As already introduced in section \ref{generation}, we will focus here only on the nine structures that are UFDs.
Multiple scattering effects are studied by analyzing the properties of the Green's matrix that is defined as:
\begin{equation}\label{Green}
G_{ij}=i\left(\delta_{ij}+\tilde{G}_{ij}\right)
\end{equation}
where the elements $\tilde{G}_{ij}$ are given by \cite{RusekPRE2D}:
\begin{eqnarray}\label{GreenOur}
\begin{aligned}
\tilde{G}_{ij}=\frac{i}{4}H_0(k_0|\textbf{r}_i-\textbf{r}_j|)
\end{aligned}
\end{eqnarray}
with $H_0(k_0|\textbf{r}_i-\textbf{r}_j|)$ the zero-order Hankel function of the first kind, $k_0$ the wavevector of light, and $\textbf{r}_i$ the position of the $i$-th scatterer in the array. The non-Hermitian matrix (\ref{Green}) describes the electromagnetic coupling among the dipoles and it enables the systematic study of the scattering properties of two-dimensional (2D) waves with an electric field parallel to the invariance axis of the CPAs (i.e., $z$-field component) \cite{Leseur}. Although the 2D model defined by the matrix (\ref{Green}) does not take into account the vector nature of light \cite{SgrignuoliVogel,SgrignuoliDiffusive,SkipetrovPRL}, it still provides robust physical information on light localization in complex 2D media \cite{RusekPRE2D,Lagendijk,sgrignuoli2021hyperuniformity}. Moreover, this 2D model  has been recently employed to describe the ``disorder-induced" transparency in high-density hyperuniform materials \cite{Leseur}, to correctly describes the coupling between two-level atoms in a structured reservoir \cite{Caze,Bouchet}, and to design aperiodic arrays for the efficient generation of two-photon processes \cite{TPSE}. In addition, it can
also be conveniently utilized as the starting point for the design
of three-dimensional photonic devices based on the membrane geometry \cite{Trojak1,Trojak2}.
We will study the Thouless number $g$ and the level spacing statistics in the low and high optical density $\rho\lambda^2$ regimes by numerically diagonalizing the $N\times N$ matrix\,(\ref{GreenOur}), where $\rho$ is the number of scatterers per unit area and $\lambda$ is the optical wavelength. In the following, we will focus on a subset of four representative arrays that fully capture the main features of the wave transport properties of CPAs. In particular, we will discuss the CPAs in the associated quadratic fields $\mathbb{Q}(\sqrt{-2})$,  $\mathbb{Q}(\sqrt{-11})$, $\mathbb{Q}(\sqrt{-67})$, and $\mathbb{Q}(\sqrt{-163})$. The  results obtained from the diagonalization of the Green's matrix for all the other structures are presented in Figures\,S1 and S2 of the Supplementary Materials. 

We find that at low optical density (i.e., $\rho\lambda^2=10^{-6}$), all the investigated systems are in the diffusive regime. In fact, their eigenvalue distributions, color-coded according to the $\log_{10}$ of the modal spatial extent (MSE), do not show the formation of any long-lived scattering resonances. The MSE parameter quantifies the spatial extension of a given scattering resonances $\Psi_i$ of the system and it is defined as \cite{SgrignuoliACS}:
\begin{equation}
\text{MSE}=\left(\displaystyle\sum\limits_{i=1}^{3N} \left|\Psi_i\right|^2\right)^2\Big/\displaystyle\sum\limits_{i=1}^{3N}  \left|\Psi_i\right|^4
\end{equation}
To gain more insight on the transport properties of CPAs, we study the behavior of the Thouless number $g$ as a function of the frequency $\omega$ \cite{SgrignuoliVogel,SgrignuoliDiffusive,sgrignuoli2021hyperuniformity}. Specifically, we sample the real parts of the eigenvalues $\Lambda_n$ in 500 equi-spaced intervals and we calculate the Thouless number in each frequency sub-interval by using the relation  \cite{SgrignuoliVogel,SgrignuoliDiffusive}:    
\begin{equation}\label{Thouless}
g(\omega)=\frac{\overline{\delta\omega}}{\overline{\Delta\omega}}=\frac{(\overline{1/\Im[\Lambda_n]})^{-1}}{\overline{\Re[\Lambda_n]-\Re[\Lambda_{n-1}]}}
\end{equation}
The symbol $\overline{\{\cdots\}}$ in eq.\,(\ref{Thouless}) denotes the sub-interval averaging operation, while $\omega$ indicates the central frequency of each sub-interval. We have verified that the utilized frequency sampling resolution does not affect the presented results. The outcomes of this analysis are reported in Figure\,\ref{FigLowOD}\,(e-h). Consistently with the low value of the optical density, we found that the Thouless number is always larger than the one, which corresponds to the diffusion regime.
\begin{figure*}[t!]
\begin{center}
\includegraphics[width=0.87\textwidth]{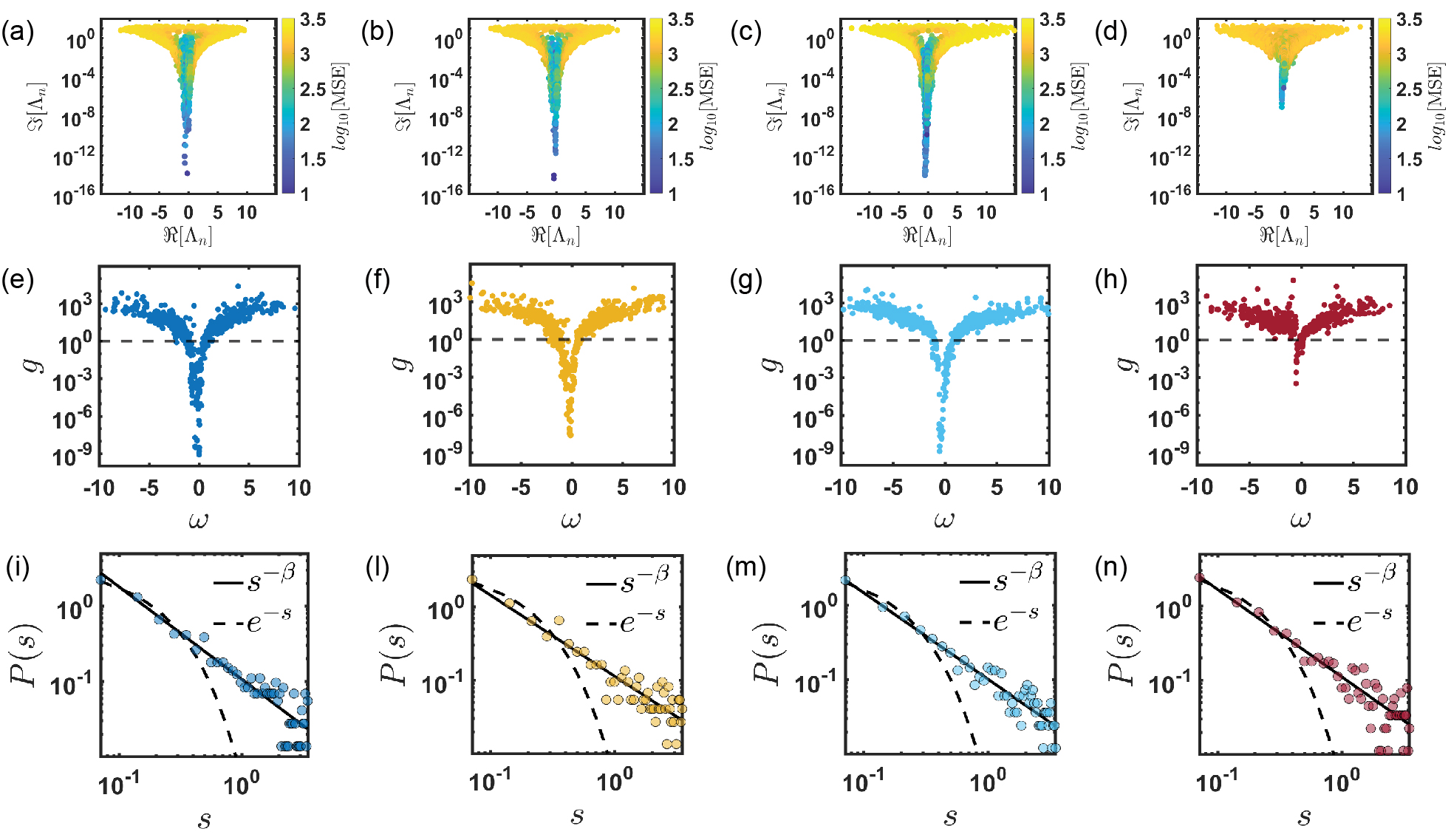}
\end{center}
\caption{Complex eigenvalue distributions, Thouless number, and level spacing statistics of the same structures analyzed in Figure\,\ref{FigLowOD} are reported, respectively, in panels (a-d), (e-h), and (i-n) when $\rho\lambda^2=5.5$. These data not only show a Thouless number $g<1$ at $\omega\approx0$, but also show that $P(s)$ describes level clustering with a power-law scaling (continuous black lines in panels (i-n)). The Poisson distribution that is characteristic of uniform random systems is also reported with a black dashed line as a comparison. }
\label{FigHighOD}
\end{figure*}

To corroborate our findings, we have analyzed the probability density function of the first-neighbor level spacing statistics of the complex Green's matrix eigenvalues $P(s)$,  where $s$=$|\Delta\Lambda|/\langle|\Delta\Lambda|\rangle$ is the nearest-neighbor Euclidean spacing of the complex eigenvalues $|\Delta\Lambda|$=$|\Lambda_{n+1}-\Lambda_{n}|$ normalized to its average value. The study of level statistics provides important information on the electromagnetic propagation in both closed and open scattering systems \cite{Shklovskii,Escalante}. In particular, the concept of level repulsion (i.e., $P(s)=0$ when $s$ goes to zero) is used to characterize the degree of spatial overlap between the resonant scattering modes of arbitrary systems \cite{Mehta}. In fact, the level statistics analysis allows one to unambiguously discriminate between a decloalized (diffusive) regime characterized by level repulsion and a localization regime characterized by level clustering (i.e., by the absence of level repulsion) as a function of $\rho\lambda^2$. We have shown in Ref.\cite{WangPRB} that the distribution of level spacing for the Gaussian and Eisenstein prime arrays in the low scattering regime manifests level repulsion described by the \textit{critical} cumulative probability defined as \cite{Zharekeshev}:
\begin{equation}\label{CriticalDistribution}
I(s)=\exp\left[\mu-\sqrt{\mu^2+(A_cs)^2}\right]
\end{equation} 
where $\mu$ and $A_c$ are fitting parameters. The results shown in Figure\,\ref{FigLowOD}\,(i-l) and in Figure\,S1 of the Supplementary Materials establish the critical level spacing statics as a robust property for all the other investigated CPAs as well at small optical density. The critical cumulative probability was successfully applied to describe the energy level spacing distribution of an Anderson Hamiltonian containing $10^6$ lattice sites at the critical disorder value, i.e., at the metal-insulator threshold where it is known that all the wave functions exhibit multifractal scaling properties \cite{Zharekeshev}. Our findings demonstrate the applicability of critical statistics to CPAs in the weakly scattering regime reflecting the singular-continuous nature of their diffraction spectra that support critically localized eigenmodes with self-similar scaling properties (see also Supplementary Material for more details on the singular-continuous nature of the investigated arrays).

On the other hand, at large optical density $\rho\lambda^2=5.5$, we observe the appearance of spatially confined long-lived scattering resonances, as shown in Figure\,\ref{FigHighOD}\,(a-d). Specifically, clear dispersion branches populated by scattering resonances that are localized over small clusters of dipoles are forming near $\omega\approx0$ in all the investigated CPAs (see also the Supplementary Material for additional details). Moreover, Figure\,\ref{FigHighOD}\,(a-c) show the formation of spectral gap regions where the critical scattering resonances reside, demonstrating the effect of local correlations on wave interference across the structures. Critical modes in aperiodic systems are spatially extended and long-lived resonances with spatial fluctuations at multiple length scales characterized by a power-law scaling behavior \cite{SgrignuoliMF,macia1999,RyuPRB}. Interestingly, we found that the CPA associated to $\mathbb{Q}(\sqrt{-163})$ does not support the formation of spectral gaps, reflecting the less coherent nature of interference phenomena in this aperiodic environment. Furthermore, at large optical density, we find that $g$ becomes lower than unity for $\omega\approx{0}$, indicating the onset of light localization, as reported in Figure\,\ref{FigHighOD}\,(e-h). We also observe that the non Euclidean CPA associated to the field $\mathbb{Q}(\sqrt{-163})$ features a significantly reduced number of localized scattering resonances compared to the other investigated CPAs. This behavior can be traced back to its distinctive geometrical structure characterized by the presence of polynomially generated linear clusters of primes that can only efficiently confine radiation along the vertical $y$-direction.

The presence of a clear DLT in the analyzed CPAs is also confirmed by the behavior of their level spacing statistics at large optical density. Under this condition, we observe a clear level clustering effect that is fundamentally different from the Poisson statistics predicted for homogeneous random media (black-dotted lines in Figure\,\ref{FigHighOD}\,(i-l)) and describing non-interacting, exponentially localized energy levels \cite{Mehta,Haake}. In contrast, the level spacing distributions observed in all CPAs are well-reproduced considering the inverse power law scaling curve $P(s)\sim s^{-\beta}$ that is displayed by the continuous black lines (see also Figure\,S2 of the Supplementary Materials). In the context of random matrix theory, it has been demonstrated that the power-law distribution describes complex systems with multifractal spectra that produce uncountable sets of hierarchical level clustering \cite{Cvitanovic,Geisel}. Moreover, this power-law scaling appears to universally describe the transport physics, for values of the exponent $\beta$ in the range $0.5<\beta<2$, of complex systems exhibiting anomalous diffusion. These are correlated random systems in which the width of a propagating wavepacket $\sigma^2$ increases upon propagation according to  $t^{2\nu}$ with $\nu\in[0,1]$ \cite{Cvitanovic}. Specifically, such a behavior was observed in one-dimensional scattering systems characterized by incommensurate sinusoidal modulations \cite{Geisel}, in quasi-periodic Fibonacci structures \cite{Guarneri}, in a family of tight-binding Hamiltonians defined on two-dimensional octagonal quasi-periodic tilings \cite{Benza}, and more recently, in three-dimensional scattering arrays designed from sub-random sequences  \cite{SgrignuoliDiffusive}. The exponents $\beta$ and $\nu$ are connected through the relation \cite{Cvitanovic,Geisel,DalNegroScirep}:
\begin{equation}\label{var}
 \sigma^2(t)=t^{2\nu}\sim t^{2(\beta-1)/d}
 \end{equation}
where $d$ is the system's dimensionality. By substituting the values of the parameter $\beta$  obtained from the numerical fits of the data in Figure\,\ref{FigHighOD}\,(i-l) into the eq.(\ref{var}), we find that the exponent $\nu$ is equal to 0.05 and 0.06. Similarly small values for the exponent $\nu$ are also obtained at large optical density for the non Euclidean structures, confirming the onset of the localization regime. These results indicate that a significant wave interference correction to classical diffusion can be obtained using the investigated CPAs \cite{John}. 

\begin{figure*}[t!]
\begin{center}
\includegraphics[width=0.5\textwidth]{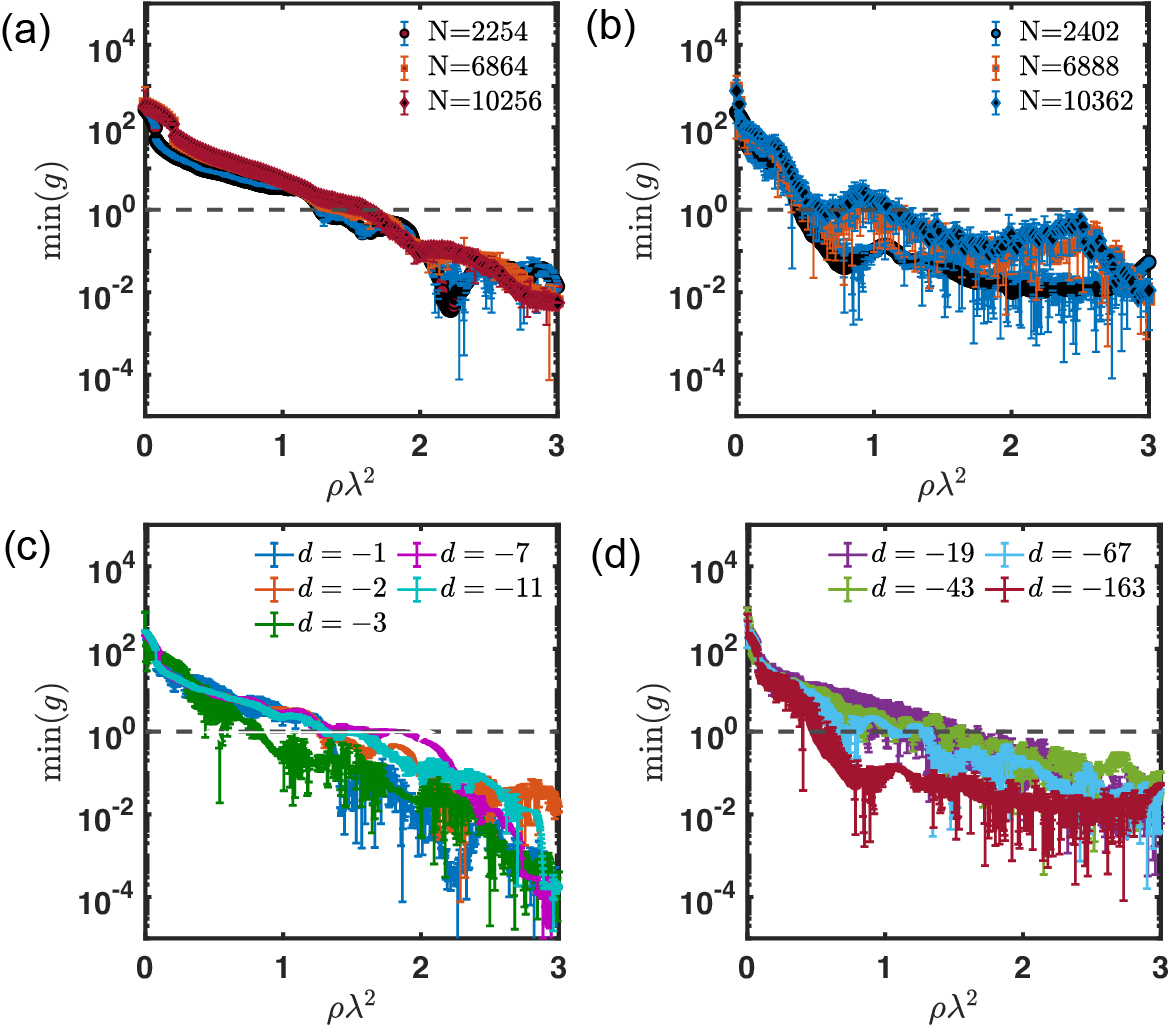}
\end{center}
\caption{Panels (a) and (b) show the scaling with respect to the system size of the minimum value of the Thouless number as a function of $\rho\lambda^2$ near by the DLT-threshold. Panel (a) refers to the CPA in the  $\mathbb{Q}(\sqrt{-2})$ field, while panel (c) to the CPA in the   $\mathbb{Q}(\sqrt{-163})$ field. Panels (c) and (d) compare the minimum value of the Thouless number as a function of $\rho\lambda^2$ for the Euclidean structures and non Euclidean structures, respectively. These data refer to systems with approximately 2500 scattering elements. The dashed line at $g=1$ marks the transition value from diffusion ($g>1$) to localization ($g<1$). The error bars are estimated from the standard deviations associated to the fluctuations of the Thouless number with respect the different choices of the binning used to sample the frequency $\omega$ in eq.\,(\ref{Thouless}).}
\label{MinThouless}
\end{figure*}
Finally, Figure\,\ref{MinThouless}\,(a-b) display the scaling, as a function of the number $N$ of prime elements, of the minimum values of the Thouless number for two representative CPAs corresponding to the quadratic fields $\mathbb{Q}(-2)$ and $\mathbb{Q}(-163)$, respectively. All these curves cross the DLT threshold value $g=1$ approximately at the same optical density, demonstrating the robustness of the transition with respect to the number of scattering dipoles in the arrays. Moreover, we show in panels (c) and (d) a comprehensive summary of the light localization properties of all the investigated CPAs, separately considering the Euclidean structures in panel (c) and the non Euclidean ones in panel (d) previously shown in Figure\,\ref{FigPointPattern}. We observe a clear DLT for all the investigated CPAs. 
To better understand the observed DLT-transition in these novel structures, we can estimate the ratio between the localization length $\xi$ and the systems size $L$. For a 2D uniform and isotropic random system, the localization length is approximately provided by \cite{Sheng}: 
\begin{equation}\label{xi}
\xi\sim l_t\exp[\pi \Re(k_e)l_t/2]
\end{equation}
with $l_t$ the transport mean free path and $\Re(k_e)$ the real part of the effective wavenumber in the medium. Although the numerical factor in eq.\,(\ref{xi}) may not be accurate \cite{Sheng,Gupta}, it nevertheless tells us that the localization length in 2D systems is an exponential function of $l_t$ and can be extremely large in the weak scattering regime (i.e., at low optical density). Moreover, in our point dipole limit the transport mean free path coincides with the scattering mean free path and can be simply estimated as $l_t=l_s=1/\rho\sigma_d$. Here, $\sigma_d$ is the cross section of a single point scatterer, which is related to its electric polarizability $\alpha(\omega)$ \cite{Caze}. At resonance, $\sigma_d$ is equal to $k_0^3|\alpha(\omega_0)|^2/4$ \cite{Lagendijk,Leseur}. Considering that, under the effective medium theory, $k_e$ can be approximated as $k_0+i/(2l_s)$ \cite{Leseur,Caze}, we can rewrite the eq.\,(\ref{xi}) as $\pi\lambda\exp[\pi^3/(2\rho\lambda^2)]/(2\rho\lambda^2)$, which directly connects the localization length of isotropic random structures with their optical density. 
In order to simply account for the discovered DLT behavior we must consider the ratio $\xi/{L}$. Consistently, we found that $\xi/L$ is very large (i.e., $\xi/L\gg1$) at low optical density indicating diffusion, while it becomes smaller than one, indicating localization, at the larger values of $\rho\lambda^2$ used in our analysis. Therefore, we have established that CPAs are promising aperiodic structures to engineer novel complex photonic environments with light localization properties. Moreover, these systems feature broadband spectra of localized optical resonances with distinctive scaling properties associated to their geometrical multifractality \cite{SgrignuoliMF,SgrignuoliRW}, as we will discuss in the next section. 

\section{Multifractality of local density of states}\label{multi}
\begin{figure*}[t!]
\begin{center}
\includegraphics[width=0.7\textwidth]{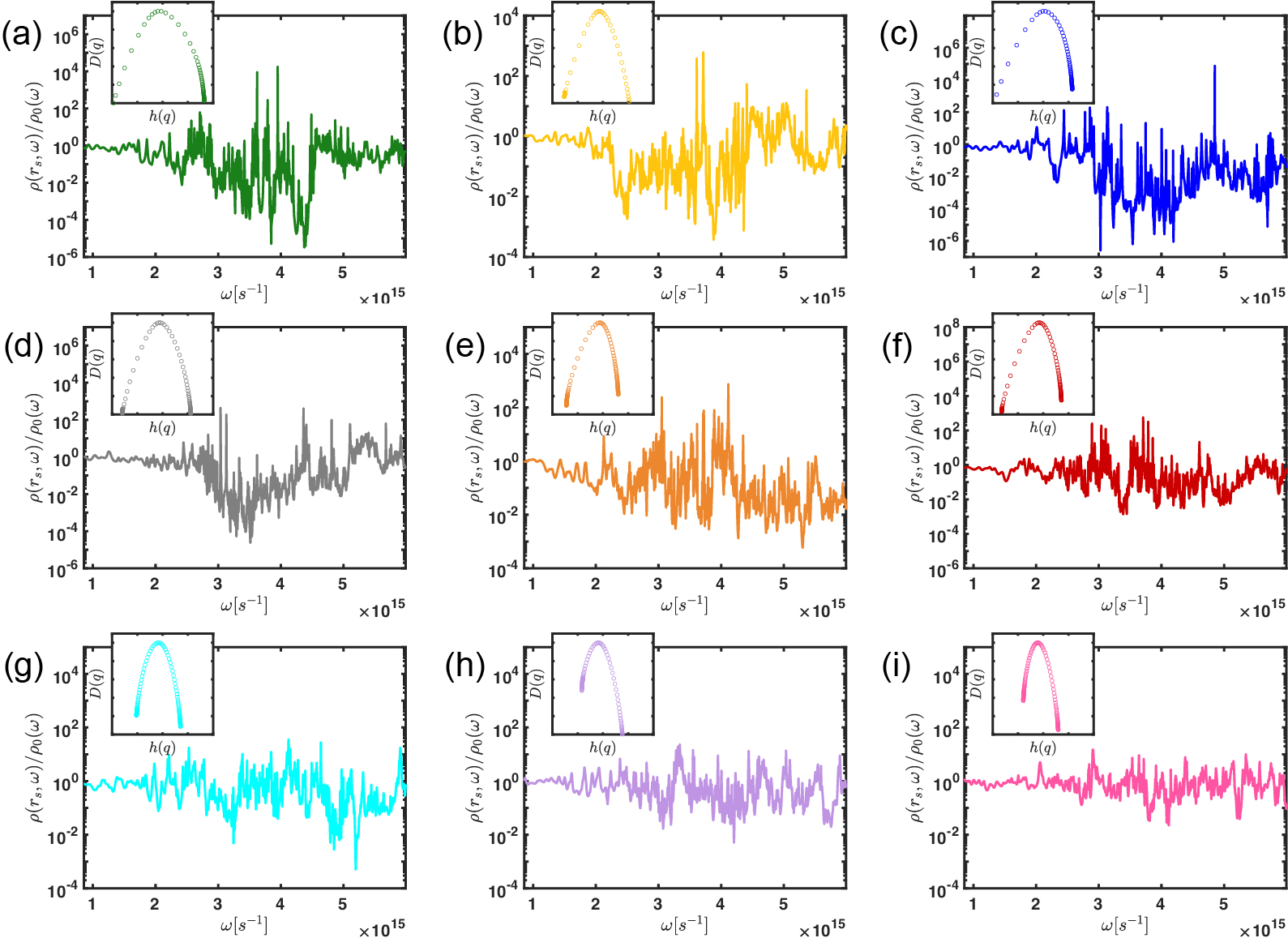}
\end{center}
\caption{(a-i) LDOS/Purcell enhancement as a function of frequency $\omega$ of vertically polarized dipoles (i.e., electric field along the z) spatially arranged according to the distributions of complex primes in the imaginary  quadratic fields shown in Figure\,\ref{FigPointPattern}. The insets show the corresponding multifractal spectra $D(q)$ as a function of $h(q)$. All the multifractal spectra are reported on the same scale with $D(q)\in[0,1]$ and $h(q)\in[0,4]$.}
\label{FigLDOSMF}
\end{figure*}

The decay dynamics of a light emitter embedded
in a complex dielectric environment can be rigorously understood by  computing its local density of states (LDOS) and the corresponding
Purcell enhancement factor as a function of frequency. The knowledge of the LDOS spectra allows one to accurately determine the spectral locations of the resonant modes with the highest quality factors, such as those that are located near the bandgap regions \cite{DalNegrocrystal2016,Trevino,TrevinoNano,DalNegrocrystal2019,DalNegroReview,Trojak1,Trojak2,SgrignuoliACS}. In order to  characterize the emission dynamics in these complex aperiodic environments, we have computed the enhancement of the LDOS with respect to its free space value, or the Purcell spectrum, for all the CPAs consisting of approximately 1000 electric dipoles with the electric polarizability $\alpha(\omega)=-4\Gamma_0c^2/[\omega_0(\omega^2-\omega_0^2+i\Gamma_0\omega^2/\omega_0)]$ \cite{Leseur}. Here, $\omega_0$ is the resonant frequency and $\Gamma_0$ identifies the linewidth. We have fixed $\omega_0=4.3\times10^{15}$ and $\Gamma_0=7.4\times10^{15}$. In this way, the polarizability $\alpha(\omega)$ describes the scattering properties of arrays composed by 30 $nm$ radius dielectric nanocylinders of constant permittivity $\varepsilon=10.5$ with an average interparticle separation of 220 $nm$ embedded in air \cite{TPSE}. In order to compute the LDOS $\rho(\textbf{r}_s,\omega)$, we have evaluated the field scattered at $\textbf{r}_s$ when the system is excited by a dipole source $p=1/\mu_0\omega^2$ also located at $\textbf{r}_s$ and oriented along the $z$-axis (i.e., parallel to the invariance axis of the scatterers). Specifically, the scattered field  is computed by solving the self-consistent Foldy-Lax equations:  
\begin{equation}\label{GreenEscat}
E_i=\mu_0\omega^2G(\textbf{r},\textbf{r}_s;\omega)p+\frac{\omega^2}{c^2}\alpha(\omega)\sum_{i\neq j}G(\textbf{r}_i,\textbf{r}_j;\omega)E_j
\end{equation}
where $\textbf{r}_i$ is the position of the scatterer $i$ and $G(\textbf{r},\textbf{r}_s;\omega)$ is the Green function\,(\ref{GreenOur}) \cite{Caze}. For a system with $N$ scattering elements, the linear system (\ref{GreenEscat}) can be solved numerically. Once the exciting field at $\textbf{r}_s$ is known, the Purcell spectrum $P(\textbf{r}_s;\omega)$ can be evaluated through the formula $4(1/4+\Im[E(\textbf{r}_s;\omega)])/(\omega^2\mu_0p)$, where $\Im[E(\textbf{r}_s;\omega)]$ is the imaginary part of the scattered field at the position of the excitation. The results of this analysis are reported in Figure\,\ref{FigLDOSMF} for all the investigated CPAs. Consistently with what previously observed for the Eisenstein and Gaussian primes \cite{TPSE}, we report for all the analyzed CPAs the presence of singular LDOS spectra with a characteristic fragmentation of spectral gap regions into smaller sub-gaps separated by localized modes. Interestingly, we discover that this fragmentation is more pronounced for the Euclidean structures compared to the non Euclidean ones, which show significantly smaller spectral gaps. The singular nature of the LDOS in complex systems with singular-continuous spectra, including the distribution of prime numbers, can be accurately described using multifractal scaling analysis \cite{Trevino,Guarneri,Wolf1,Muzy,Stanley}.

In order to characterize the  multifractal scaling of the LDOS fluctuations observed in Figure\,\ref{FigLDOSMF} for all the CPAs we apply Multifractal Detrended Fluctuation Analysis (MDFA) \cite{Kantelhardt} using the numerical routines developed by Ihlen for the study of non-stationary complex signals \cite{Ihlen}.
The MDFA is a powerful technique that extends the traditional Detrended Fluctuation Analysis (DFA) \cite{peng1994mosaic} to the case of non-stationary time series with multifractal scaling properties. This is achieved by  considering the scaling of their fluctuations with respect to polynomial trends defined over sub-intervals of the analyzed signals (i.e., local detrending). 
The local nature of the scaling procedure is essential because, in contrast to homogeneous fractals (or monofractals), the scaling of multifractals is locally defined around each point. More precisely, the scaling behavior around any point of a general multifractal measure $\mu$ is accounted by its local (i.e., position dependent) power law $\mu({\vec {x}}+{\vec {a}})-\mu({\vec {x}})\sim a^{h({\vec {x}})}$. Here the  generalized Hurst exponent $h({\vec {x}})$, or singularity exponent, quantifies the strength of the singularity of the multifractal signal around that specific point. In addition, it can be shown that  the set of all the points that share the same singularity exponent is a fractal set with a continuous distribution of fractal dimensions, characterized by the singularity (or multifractal) spectrum \cite{falconer2004fractal}. 
The MDFA enables accurate determination of the multifractal parameters of a signal, including the multifractal spectrum. This is achieved by considering the local scaling of generalized fluctuations with respect to smooth trends over piecewise sequences of locally approximating polynomial fits, i.e., ${\displaystyle F_{q}(n)\propto n^{h(q)}}$ where $h(q)$ is the generalized Hurst exponent, or $q$-order singularity exponent \cite{Ihlen}. The generalized parameter $h(q)$ reduces to the conventional Hurst exponent $H\in[0,1]$ for stationary signals.
\begin{figure*}[t!]
\begin{center}
\includegraphics[width=0.7\textwidth]{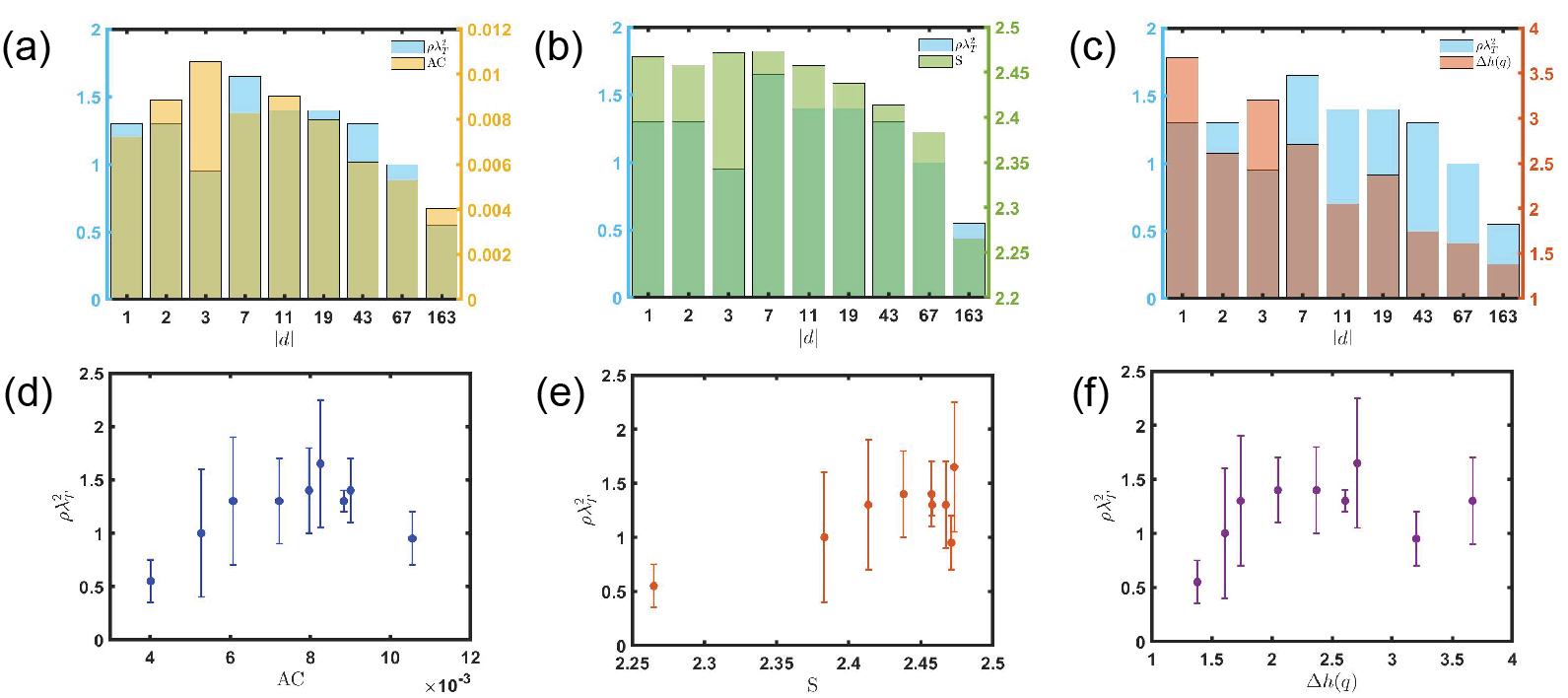}
\end{center}
\caption{Histograms of the estimated values of the DLT-threshold ($\rho\lambda^2_T$) of all the investigated CPAs together with their algebraic connectivity $AC$ in panel (a), nodal entropy $S$ in panel (b), and multifractal width $\Delta h(q)$ in panel (c). Panels (a-c) compare $\rho\lambda^2_T$ with respect to $AC$, $S$, and $\Delta h(q)$, respectively.}
\label{FigHisto}
\end{figure*}
In our analysis the multifractal LDOS signal $x_{t}$ is regarded as a discrete series of data points labelled by the integer parameter ${t}\in\mathbb{N}$ and the generalized fluctuations $F_{q}(n)$ are defined as the $q$-order moments over $N$ intervals of size $n$ according to \cite{Kantelhardt,Ihlen}:
\begin{equation}\label{moment}
{\displaystyle F_{q}(n)=\left({\frac {1}{N}}\sum _{t=1}^{N}\left(X_{t}-Y_{t}\right)^{q}\right)^{1/q}}
\end{equation}
where:
\begin{equation}
X_t=\sum_{k=1}^t (x_k-\langle x\rangle)    
\end{equation}
Moreover, ${\displaystyle X_{t}}$ is subdivided into time windows of size ${\displaystyle n}$ samples and ${\displaystyle Y_{t}}$ denotes the piecewise sequence of approximating trends obtained by local least squares fits. Finally, ${\displaystyle \langle x\rangle }$ is the mean value of the analyzed time series corresponding to the LDOS signal. Finally, the multifractal spectra of the LDOS, shown in the insets of Figure\,\ref{FigLDOSMF} for the corresponding CPA, are computed from the mass exponent $\tau(q)$ using the Legendre transform \cite{Kantelhardt,TPSE}:
\begin{equation}
D(q)={q}\tau^{\prime}(q)-\tau(q)  
\end{equation}
The broad and downward concavities of the multifractals spectra of the LDOS of CPAs are indicative of the strong multifractal behavior of these complex systems \cite{SgrignuoliMF,TPSE}. The width $\Delta{h(q)}$ of the support of the multifractal spectra is directly related to the degree of spatial non-uniformity of the corresponding signals \cite{SgrignuoliMF}. Our results demonstrate that the strongest LDOS multifractality occurs for the Eisenstein prime structures and it drops significantly for the more non Eucliedean CPAs that are characterized by more regular geometries. This trend is consistent with the reduction in nodal entropy and algebraic connectivity that was previously identified for non Eucliedean CPAs. 

We are now well-position to  establish robust structure-property relationships valid for all the investigated CPAs by correlating their structural and spectral information.
In particular, here we investigate the connection between geometrical and  transport properties by showing in Figure\,\ref{FigHisto} the dependence of the critical optical density $\rho\lambda^2_{T}$ at which the DLT appears in the different structures on the AC parameter, the node entropy, and the width of the multifractal spectra $\Delta h(q)$.   Moreover, we compare in panels (a), (b), and (c) the histogram plots of $\rho\lambda^2_{T}$ (blue bars) as a function of the absolute value of the radicand of the field $|d|$ with the histograms plots of the AC density (yellow bars), the node entropy $S$ (green bars), and the width $\Delta h(q)$ (orange bars) of the multifractal LDOS/Purcell spectra, respectively. Interestingly, we observe that the AC and nodal entropy strongly correlate with the DLT threshold values for all the systems. In particular, these parameters decrease when $|d|$ increases and this correlation is particularly evident in the non Euclidean CPAs. This analysis shows that non Euclidean CPAs are structurally more homogeneous than their Euclidean counterparts,  consistently with their reduced nodal entropy.  The implications of these structural properties on transport are displayed clearly in panels (d), (e), and (f), where we report $\rho\lambda^2_{T}$ as a function of the structural parameters $AC$, $S$, and $\Delta h(q)$, respectively. The error bars in these figures reflect the uncertainty in the estimation of the $\rho\lambda^2_{T}$ at which the minimum values of $g$ becomes less than unity due to the fluctuations of the Thouless number with respect to size of the binning interval used to compute eq.\,(\ref{Thouless}). Notably, we found an approximate linear relationship with respect to the nodal entropy $S$ (see Figure\,\ref{FigHisto}\,(e)). Our findings demonstrate that the structural complexity and inhomogeneity of CPAs are the key ingredients that drive the discovered DLT transitions. These results establish relevant structure-property relationships that qualitatively capture the complex interplay between light scattering and localization in these novel complex environments. Building on these structurally-driven transport properties, in the final section of the paper we will discuss the ability of CPAs to enter the strong coupling regime that is relevant for the engineering of novel quantum sources.

\section{Prime numbers in the quantum regime: Rabi splitting}
Spectral fractality has profound implications on the relaxation dynamics of quantum sources \cite{akkermans2013spontaneous,TPSE}. In the following section, we will address the decay rate of a two-level atom (TLA) with transition frequency $\omega_{if}=\hat{\omega}$ embedded inside a CPA-structured vacuum with multifractal scaling properties. In particular, we will focus on the strong coupling regime where the interaction of the TLA with CPA photonic environment is faster than the dissipation of energy from either system. The strong coupling is fundamental in quantum-based measurements and quantum information protocols, as well in the testing of the long-standing questions about macroscopic quantum coherence \cite{braginsky1995quantum,mancini1997ponderomotive,hennessy2007quantum,yoshie2004vacuum}.

As a proof-of-concept demonstrator, we focus here on the radiation from Eisenstein prime arrays. These structures are the best candidate ones to achieve strong light-matter interaction. In fact, based on our previous analysis, the Eisenstein prime arrays possess the most compact (high-density) and complex aperiodic geometry characterized by the highest nodal entropy value and the broadest multifractal LDOS spectrum. In addition, among all the CPAs the Eisenstein structures also display the highest degree of rotational symmetry, which favors the formation of spectral badgap regions in low-index photonic environments  \cite{Pollard} and, in combination with aperiodic order, is known to give rise to highly-localized optical modes \cite{SgrignuoliVogel,Trojak1,Trojak2}. This feature is clearly visible in Figure\,\ref{FigRS}\,(a), where we show the Purcell spectrum $P(\textbf{r}_s;\omega)=\rho(\textbf{r}_s,\omega)/\rho_0(\omega)$ evaluated at the center of an Eisenstein array with $N=1062$ scattering particles. A broadband distribution of highly-confined band-edge resonant modes can be observed in the Purcell spectrum of this structure. An enlarged view of the highest peak of $P(\textbf{r}_s;\omega)$ is shown along with its Lorentzian fit in Figure\,\ref{FigRS}\,(b). This localized optical mode, shown in the inset of panel (b), has an eigenfrequency $\omega=4.85\times10^{15}s^{-1}$, and an effective linewidth $\Gamma=8.8\times10^{9}s^{-1}$, yielding a quality factor $Q=\omega/\Gamma=1.5\times10^5$ and a Purcell enhancement factor $P=5.5\times10^5$. The inset of panel (b) is computed using a square grid of vertically-polarized electric dipoles with a resolution of 19$nm$ (i.e., almost $5\times10^5$ dipoles are considered). We now use this highly-localized mode to demonstrate numerically the strong coupling regime following the approach described in references \cite{Leseur,Caze}. The computational analysis proceeds as follows: first, we add an extra scattering dipole at position $\textbf{r}_s$ with the electric polarizability:  
\begin{equation}\label{TLApol}
\alpha_{TLA}(\omega)=-\frac{4\Gamma_{r} c^2}{\hat{\omega}[(\omega^2-\hat{\omega}^2+i(\Gamma_{r}+\Gamma_{nr})\omega^2/\hat{\omega})]}
\end{equation}
The parameters $\Gamma_{r}$, and $\Gamma_{nr}$ are, respectively, the radiative and intrinsic nonradiative linewidth \cite{Leseur,Caze}. 
We then tune its resonance frequency $\hat{\omega}$ to the one of the localized mode (i.e., $\hat{\omega}=\omega_C$). For simplicity, we assume $\Gamma_{nr}=0$. As a consequence, the electric polarizability (\ref{TLApol}) describes either a classical resonant scatterer or a quantum two-level atom far from saturation \cite{Bouchet}. Subsequently, we illuminate the hybrid system composed of the ``CPA+TLA" with an incident plane wave, resulting in the coupled self-consistent equation \cite{Caze}:    
\begin{figure*}[t!]
\begin{center}
\includegraphics[width=0.6\textwidth]{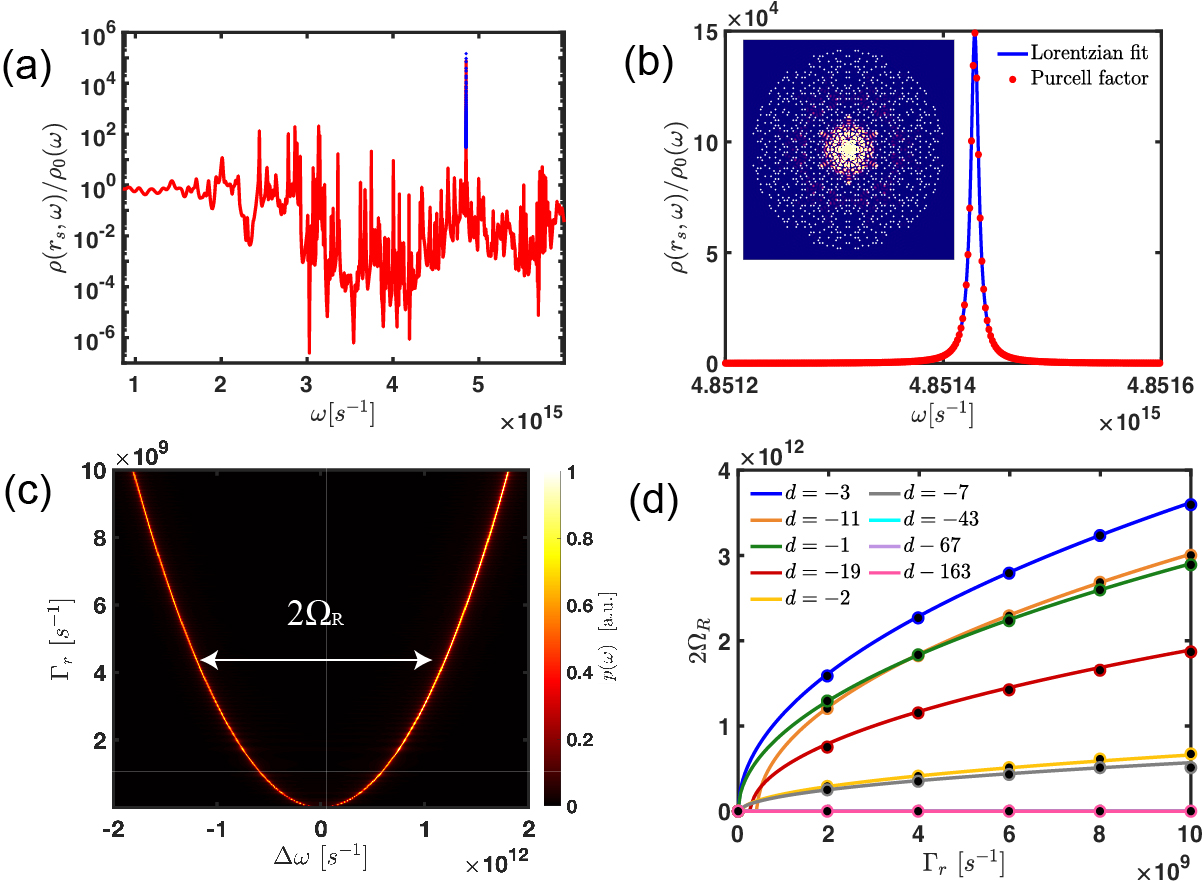}
\end{center}
\caption{(a) LDOS enhancement/Purcell factor as a function of frequency $\omega$ for vertically polarized  electric dipoles spatially according to  the Eisenstein prime arrays evaluated at the center of the structure. The blue markers identify the optical mode with the largest Purcell enhancement that is fitted by a Lorentzian lineshape, as reported in panel (b). Panel (c) displays the strong coupling between a TLA, positioned at the center of the structure, and the optical mode reported in panel (b). The Rabi splitting is evaluated numerically by calculating the absolute value of the induced dipole moment $|p(\omega)|$ of the TLA as  a function of the detuned frequency $\Delta\omega$ and the radiative linewidth $\Gamma_r$. (d) Rabi splitting $2\Omega_R$ of a TLA coupled with the optical modes with the largest Purcell enhancement factor for each investigated CPA. The continuous lines are the analytical predictions obtained from the analysis of the complex poles of the dressed polarizability while the markers indicate the numerical data.}
\label{FigRS}
\end{figure*}

\begin{equation}\label{eq13caze}
E_i=E_0(\mathbf{r}_i,\omega)+k^2\alpha(\omega)\sum_{i\neq j} G_0(\mathbf{r}_i,\mathbf{r}_j,\omega)E_j+k^2\alpha_{TLA}(\omega)G_0(\mathbf{r}_i,\mathbf{r}_s,\omega)E_{TLA}
\end{equation}
where the locally exciting field on the TLA is given by:
\begin{equation}\label{eq14caze}
E_{TLA}=E_0(\mathbf{r}_s,\omega)+k^2\alpha(\omega)\sum_{i=1}^N G_0(\mathbf{r}_s,\mathbf{r}_i,\omega)E_i
\end{equation}
By substituting eq.(\ref{eq14caze}) into eq.(\ref{eq13caze}), we can evaluate the exciting field on each of the $N$ scatterers in the CPA dielectric environment by solving the self-consistent Foldy-Lax equations:
\begin{equation}\label{Alternativosistem}
\begin{bmatrix} 
E_1^{ext}\\
\\
E_2^{ext}\\
\\
\vdots\\
\\
E_N^{ext}\\
\end{bmatrix} 
= \begin{bmatrix} 
E_1^{0}\\
\\
E_2^{0}\\
\\
\vdots\\
\\
E_N^{0}\\
\end{bmatrix} \\
+
\textbf{M} 
\begin{bmatrix} 
E_1^{ext}\\
\\
E_2^{ext}\\
\\
\vdots\\
\\
E_N^{ext}\\
\end{bmatrix} 
\end{equation}
where \textbf{M} is the $N\times N$ matrix defined by:
\begin{equation}
\begin{bmatrix} 
k^4\alpha_{TLA}\alpha G_0^{1s}G_0^{s1}&k^2\alpha G_0^{12}+k^4\alpha_{TLA}\alpha G_0^{1s}G_0^{s2}&\dots&k^2\alpha G_0^{1N}+k^4\alpha_{TLA}\alpha G_0^{1s}G_0^{sN}\\
\\
k^2\alpha G_0^{21}+k^4\alpha_{TLA}\alpha G_0^{2s}G_0^{s1}&k^4\alpha_{TLA}\alpha G_0^{2s}G_0^{s2}&\dots&k^2\alpha G_0^{2N}+k^4\alpha_{TLA}\alpha G_0^{2s}G_0^{sN}\\
\vdots &\vdots&\ddots& \vdots&\\
\\
k^2\alpha G_0^{N1}+k^4\alpha_{TLA}\alpha G_0^{Ns}G_0^{s1}&k^2\alpha G_0^{N2}+k^4\alpha_{TLA}\alpha G_0^{Ns}G_0^{s2}&\dots&k^4\alpha_{TLA}\alpha G_0^{Ns}G_0^{sN}\\
\end{bmatrix}
\end{equation}
Once the locally exciting fields on each scattering dipole are known, we can compute the induced dipole moment of the probe scatterer $p_{TLA}=\epsilon_0\alpha_{TLA}(\omega)E_{TLA}(\textbf{r}_s,\omega)$, whose absolute value yields the Rabi splitting \cite{Leseur,Caze}. Figure\,\ref{FigRS}\,(c) shows the resulting spectrum as a function of $\Delta\omega$ and the radiative linewidth $\Gamma_r$. We note that the single Lorentzian peak is now  split into two symmetric peaks by increasing $\Gamma_r$, which demonstrates the onset of Rabi-splitting in the system. 

To provide a comprehensive overview of the strong coupling regime of CPA dielectric environments, we have extended our numerical study to all the investigated CPAs. The results of this analysis are presented in Figure\,\ref{FigRS}\,(d). The markers denote the numerical results evaluated by solving eq.\,(\ref{Alternativosistem}), while the continuous lines refer to the analytical expression of the Rabi frequency $\Omega_R$ derived by analyzing the poles of the dressed polarizability of the system $\hat{\alpha}(\omega)=\alpha_{TLA}/[\mathbf{I}-k^2\alpha_{TLA}(\omega)\mathbf{S}^{reg}(\mathbf{r}_s,\mathbf{r}_s,\omega)]$ \cite{Caze,Bouchet}. Here, $\mathbf{S}^{reg}$ is the regularized scattered Green function \cite{Caze,Bouchet}. $\Omega_R$ is equal to $(g^2-\Gamma_r^2/16)^{1/2}$, where the coupling constant $g$ is given by the relation $\sqrt{(\Gamma_r\Gamma_C P_C/4)}$. To  evaluate the Rabi-frequency as a function of $\Gamma_r$, we have selected the optical mode that produces the largest Purcell enhancement in each CPAs. Figure\,\ref{FigRS}\,(d) summarizes our results that determine the ability of the investigated CPAs to produce strongly-coupled hybrid states. Interestingly, we found that, with the exception of the CPA corresponding to $d=-19$, all the non Euclidean structures do not produce any Rabi-splitting, reflecting their inability to support pronounced bandgap regions with highly-localized band-edge modes.The anomalous case of $d=-19$ case, being structurally very similar to the Euclidean CPAs (see Table 1 and Figure\,\ref{FigHisto}) shares similar performances and represents the transition boundary between the two different classes.  All the remaining non Euclidean CPAs show significantly reduced density, algebraic connectivity, and node entropy resulting in ``less disordered" dielectric environments, in agreement with the smaller widths of their $P(\textbf{r}_e,\omega)$ multifractal spectra. In contrast, the ``more disordered" geometries of the Euclidean CPAs give rise to more suitable photonic environments for the generation of highly-localized optical resonances in the strong coupling regime.       

\section{Conclusions}
In conclusion, we have presented a comprehensive analysis of the structural, spectral, and localization properties of novel aperiodic arrays of scattering dipoles that inherit the intrinsic complexity of prime numbers in imaginary quadratic fields. We have presented a theoretical analysis that combines the interdisciplinary methods of spatial statistics and graph theory analysis of point patterns with the rigorous Green's matrix spectral approach to the multiple scattering problem of optical waves. Our work has unveiled the relevant structural properties that result in a wave localization transition in the investigated structures. 
Specifically, we have demonstrated the onset of a Delocalization-Localization Transition (DLT) by a comprehensive analysis of the spectral properties of the Green's matrix and the Thouless number of these systems as a function of the optical density. Moroever, we showed that CPAs are gapped systems with a multifractal spectrum of localized resonances resulting in a far-richer scattering and localization behavior compared to periodic and uniform random structures. Furthermore, we establish a connection between their localization, multifractality, and graph connectivity properties.
Finally, we employed a semi-classical approach to demonstrate and characterize the strong coupling regime of quantum emitters embedded in these novel aperiodic environments. This study provides access to engineering design rules for the fabrication of novel and more efficient classical and quantum sources as well as photonic devices with enhanced light-matter interaction based on the intrinsic structural complexity of prime numbers in imaginary quadratic fields.
Our comprehensive analysis is meant to stimulate the development of novel design principles to achieve broadband enhancement of light-matter interactions in both the classical and quantum regimes.

\section*{Author Contributions}
L.D.N. conceived, lead, supervised the research activities, and wrote the manuscript with critical feedback from all the authors. All the authors performed numerical computations, analyzed, and organized the final results. All authors contributed to discussions and manuscript revisions.
\begin{acknowledgments}
 L.D.N. acknowledges the support from the Army Research Laboratory under Cooperative Agreement Number W911NF-12-2-0023.
\end{acknowledgments}

\end{document}